\documentclass[preprint,journal]{vgtc}
\ieeedoi{10.1109/TVCG.2019.2934415}

\ifpdf
  \pdfoutput=1\relax                   
  \usepackage{graphicx}                
  \DeclareGraphicsExtensions{.pdf,.png,.jpg,.jpeg} 
\else
  \usepackage{graphicx}                
  \DeclareGraphicsExtensions{.eps}     
\fi%

\graphicspath{{figures/}{pictures/}{images/}{plots/}{./}} 

\usepackage{microtype}                 
\PassOptionsToPackage{warn}{textcomp}  
\usepackage{textcomp}                  
\usepackage{mathptmx}                  
\usepackage{times}                     
\usepackage{cite}                      
\usepackage{tabu}                      
\usepackage{booktabs}                  

\usepackage[utf8]{inputenc}
\usepackage{array}
\newcolumntype{L}[1]{>{\raggedright\let\newline\\\arraybackslash\hspace{0pt}}m{#1}}
\newcolumntype{C}[1]{>{\centering\let\newline\\\arraybackslash\hspace{0pt}}m{#1}}
\newcolumntype{R}[1]{>{\raggedleft\let\newline\\\arraybackslash\hspace{0pt}}m{#1}}
\usepackage{textcomp}
\usepackage{xcolor,colortbl}
\usepackage{enumitem}
\definecolor{newcontent}{HTML}{000000}
\definecolor{newcontent2}{HTML}{000000}
\definecolor{highlight}{HTML}{FF0000}
\definecolor{updates}{HTML}{FF0000}
\definecolor{greenhighlight}{HTML}{00CC00}

\newcommand{\highlight}[1]{\color{black}#1 \color{black}}
\newcommand{\update}[1]{\color{black}#1 \color{black}}

\onlineid{1013}

\vgtccategory{Research}
\vgtcpapertype{Empirical Study}

\title{Evaluating an Immersive Space-Time Cube\\Geovisualization for Intuitive Trajectory Data Exploration}

\author{Jorge A. Wagner Filho, \textit{Student Member, IEEE}, Wolfgang Stuerzlinger, \textit{Member, IEEE}, \\and Luciana Nedel, \textit{Member, IEEE}}
\authorfooter{
\item
 Jorge A. Wagner Filho is with the Federal University of Rio Grande do Sul and Simon Fraser University. E-mail: jawfilho@inf.ufrgs.br.
\item
 Wolfgang Stuerzlinger is with Simon Fraser University. E-mail: w.s@sfu.ca.
\item
 Luciana Nedel is with the Federal University of Rio Grande do Sul. E-mail: nedel@inf.ufrgs.br.
}

\shortauthortitle{Wagner Filho \MakeLowercase{\textit{et al.}}: Immersive Space-Time Cube Geovisualization}

\abstract{A Space-Time Cube enables analysts to clearly observe spatio-temporal features in movement trajectory datasets in geovisualization. However, its general usability 
is impacted by a lack of depth cues, a reported steep learning curve, and the requirement for efficient 3D navigation. 
In this work, we investigate a Space-Time Cube in the Immersive Analytics domain. 
Based on a review of previous work and selecting an appropriate exploration metaphor, we built a prototype environment where the cube is coupled to a virtual representation of the analyst's real desk, and zooming and panning in space and time are intuitively controlled using mid-air gestures. We compared our immersive environment to a desktop-based implementation in a user study with 20 participants across 7 tasks of varying difficulty, which targeted different user interface features. 
To investigate how performance is affected in the presence of clutter, we explored two scenarios with different numbers of trajectories. 
While the quantitative performance was similar for the majority of tasks, large differences appear when we analyze the patterns of interaction and consider subjective metrics. The immersive version of the Space-Time Cube received higher usability scores, much higher user preference, and was rated to have a lower mental workload, without causing participants discomfort in 25-minute-long VR sessions.%
} 


\CCScatlist{ 
 \CCScat{H.5.1}{Information Interfaces and Presentation}{Multimedia Information Systems}{Artificial, augmented, and virtual realities}
}

\teaser{
  \centering
  \includegraphics[width=\linewidth]{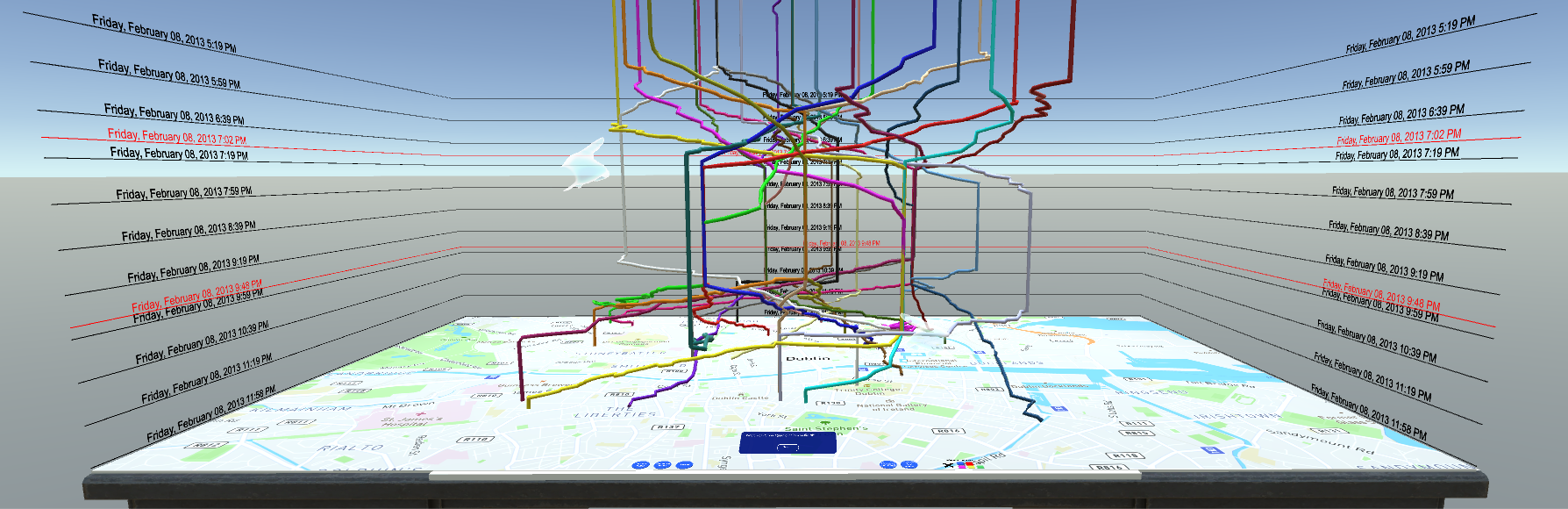}
  \caption{Our immersive space-time cube (STC) aims to lower the known steep learning curve of this three-dimensional spatio-temporal representation. The base map is coupled to a virtual reproduction of the analyst's own real desk and all data manipulations and queries are implemented through intuitive gestures and tangible controls, increasing usability and comfort, while decreasing mental workload.}
	\label{fig:teaser}
}

\vgtcinsertpkg

\begin{document}


\firstsection{Introduction}

\maketitle

Visualizing how objects' or people's positions vary over time and facilitating the extraction of meaningful patterns and 
insights from such visualizations is a topic of growing interest for a wide range of stakeholders, from regular people planning their journeys to human geographers and decision-makers preparing for possible situations in the future 
\textcolor{black}{or analyzing the past}~\cite{andrienko2010space}. 
Nowadays, GPS-enabled mobile devices, telecommunication networks \cite{calabrese2015urban}, and even social networks \cite{noulas2011empirical} can easily collect large and detailed datasets recording people's positions over time, with applications in a broad range of domains, including urban planning, mobility optimization, sports analytics, and behavioral studies. 

Due to their spatial nature,
such trajectory datasets are typically explored with the aid of 2D map representations via animations to represent variations across time, but this often obscures 
relevant temporal features and patterns in the data. 
An alternative that aims to more adequately address the temporal component is the Space-Time Cube (STC), a 3D representation that uses an additional, third 
dimension to represent time. Originally proposed in the 1970s by Hägerstrand \cite{hagerstraand1970people}, it attracted more attention after the introduction of computer-based systems that enable an interactive analysis from different points of view \cite{kraak2003space}, essential 
for its effective use. 
The STC enables a clear observation of various features in time and space, such as varying movement durations and speeds (shown as non-vertical lines lengths and slopes), stop durations and locations (which are represented by a vertical line) and meetings between different individuals (when they share the same position in time and space). Such analysis is usually (much) more difficult with standard 2D map-based visualizations.

However, and similar to other 3D data representations, the STC suffers from well-known limitations in terms of perception and interaction when it is used in conventional desktop setups \cite{Ware:2004:IVP}. These are a result of the difficulties in estimating distances and visual depths from monocular depth cues alone and the mismatch in terms of control between a 3D environment and a 2D interaction device such as the mouse, \update{as well as of the challenges introduced by occlusion and visual clutter.} This is even worse when taking into account that domain experts are typically not---and should not be required to be---trained in 3D manipulation. Previous work identified that experts specifically complained about a steep learning curve for the STC 
\cite{kveladze2015space}. 

In this work, we hypothesize that the usability of STCs could greatly benefit from an immersive approach. Immersive Analytics applications combine stereoscopic displays and intuitive 3D interaction techniques to better explore 3D data representations \cite{chandler2015immersive, immersiveanalyticsbook}, with promising results found in multiple domains \update{\cite{kwon2016study,WagnerFilho2018VR,coffey2011slice,drouhard2015immersive}.} 
\update{The ability to easily and quickly change the viewpoint through smaller or larger head movements enables users to deal with occlusion and visual clutter effectively.
Also, head movements combined with stereopsis facilitate the comparison of distances and depths \cite{WagnerFilho2018VR}, as well as the understanding of spatial relationships.} 
Yet, so far little work has explored immersive approaches for spatio-temporal data. \highlight{Even though prior work failed to identify benefits of stereopsis for the exploration of STCs, it 
only considered event data instead of complete 3D trajectories, and a very limited set of tasks \cite{kjellin2010different}. We believe that a broader 
set of tasks, the integration of new interaction methods, and the 
inclusion of qualitative factors justify a revisitation of this matter. Moreover, we are interested in assessing not only the benefits of stereopsis but also of immersive exploration as a whole, including more direct and intuitive interaction.}

Considering the vast design space for immersive environments, it is clear that 
system design choices directly affect user experience and performance, and that user comfort, efficiency, and integrability into a traditional work environment are critical requirements for the adoption of an approach by domain experts. With this in mind, we 
reviewed various implementations of the STC in the literature (\autoref{sec:rel}) and collected the most appropriate design choices for our immersive environment (\autoref{sec:prop}).
Moreover, we decided to base our work on a virtual desk metaphor for seated exploration \cite{wagner2018virtualdesk}. 
The STC's base map is coupled to the surface of a virtual representation of the analyst's real work desk. The trajectories, rendered within arm's reach, can be intuitively manipulated with mid-air gestures and the aid of tangible control buttons positioned on the desk. 

In summary, we present the following contributions:
\begin{itemize}[noitemsep,nolistsep]
\setlength{\itemsep}{1pt}
\setlength{\parskip}{0pt}
\setlength{\parsep}{0pt}
\item A comparative user study across a large set of tasks and two data density conditions to investigate the potential of an immersive STC for spatio-temporal data exploration.
\item Design recommendations for future immersive STC systems.
\end{itemize}

In the remaining of this paper, we discuss our strategy to evaluate the new prototype implementation (\autoref{sec:howto}) and report on a controlled user study conducted with 20 \update{novice} participants (\autoref{sec:study}), comparing performance for 7 typical movement data analysis tasks in terms of subjective preferences, measured workload, accuracy and completion time, against a standard desktop-based implementation (\autoref{sec:res}). 
Moreover, we explore two different data scenarios to investigate how the presence of clutter affects each condition. We also analyzed completion times 
in terms of the usage of different interactivity features. \update{Our hypotheses are listed in \autoref{sec:study:hypos}.} 

\section{Related Work}
\label{sec:rel}

The exploration of spatio-temporal data is a frequent task in the field of geovisualization, integrating approaches from cartography, geographic information science and data visualization \cite{kraak2006visualization}. Typically such exploration involves interactive representations.
Here, we focus on the Space-Time Cube (STC) representation.

\subsection{Visualization of Movement with the Space-Time Cube}

The Space-Time Cube (STC) was originally proposed by 
Hägerstrand \cite{hagerstraand1970people} as part of his time-geography studies, which saw space and time as inseparable. The cube's main idea is to use the axis perpendicular to a map's surface to represent time, thus allowing the precise display of trajectories (called Space-Time Paths), which makes movement features, such as varying speeds, meetings between people, and stationary periods (clearly represented as vertical lines, called Stations) more evident. 
Since the original STC representation required manual redrawing for the inspection of alternative points of view, its application remained limited for a long time. However, with the advent of interactive computer-based systems, the STC was revisited by Kraak~\cite{kraak2003space} as part of an extended visualization environment for movement data. A similar concept was also adopted by \textit{GeoTime} \cite{geotime2004}.

Since then, interactive 
STCs have been employed to visualize movement trajectories in various domains, such as vessel navigation \cite{willems2011evaluation}, running sports \cite{kraak2008geovisualization}, historic events \cite{kraak2008geovisualization}, and urban mobility \cite{amini2015impact,kveladze2017researching,gonccalves2015cartographic}. 
The STC can also be applied in the exploration of other kinds of spatio-temporal datasets, such as event data (e.g., occurrences of earthquakes \cite{gatalsky2004interactive}, crimes \cite{morgan2010visual} or disease cases \cite{kraak2008geovisualization,kjellin2010different}), origin-destination data (e.g.\highlight{,} movement patterns obtained from activity diaries \cite{chen2011exploratory}), and even eye-tracking recordings \cite{li2010visual}. The STC was also used by Bach et al. \cite{bach2014review} as a conceptual framework to survey temporal data visualizations.

The STC can be used either with a small set of trajectories, prioritizing the perception of individual movement behaviors, or
with a much larger set of trajectories to identify collective similarities and differences between behaviours of population subgroups, for example from different ethnic origins or socio-economic status \cite{kveladze2015space}, or between pedestrians who access a given region from different entrances \cite{kveladze2012recognition}. However, when displaying many trajectories at the same time, the STC can become
cluttered and potentially important features can be occluded, sometimes motivating the usage of aggregation techniques \cite{demvsar2010space}. 
In this work, we consider two data scenarios with varying degrees of clutter to assess the performance of immersive exploration for each case. 
In both scenarios we are mostly interested in the identification of individual movement features---which ultimately also contribute to the identification of broader patterns.
   
For the original interactive implementations of the STC, there was no clear evidence of analytical benefits~\cite{kraak2003space}. Thus, a series of follow-up studies addressed the need for controlled evaluations.

\subsection{Questions about Movement Data}
\label{sec:rel:questions}

To evaluate any movement data representation, it is important to select a set of relevant typical tasks to be conducted by data analysts in this domain. 
One of the most influential taxonomies for 
this is the \textit{Spatio-Temporal Triad} framework by Peuquet \cite{peuquet1994s}, which identifies three basic kinds of questions, corresponding to two known information components and the one being investigated: \textit{`when + where $\rightarrow$ what'}, \textit{`when + what $\rightarrow$ where'} and \textit{`where + what $\rightarrow$ when'}. Andrienko et al. \cite{andrienko2003exploratory} later revisited the notion of different \textit{reading levels} from Bertin \cite{bertin1983semiology}, indicating whether a question component (i.e., space, time, or objects) refers to a single data element (elementary question) or to a group of elements (general question). To reduce the size of the typology, however, they simplified the types of question by focusing only on whether time is given or asked: \textit{`when $\rightarrow$ where + what'} or \textit{`where + what $\rightarrow$ when'}, where each side of the arrow can be elementary or general. 
More recently, Amini et al. \cite{amini2015impact} proposed an updated taxonomy where each of the three question variables can refer either to a single value or multiple values (i.e., be it \textit{singular} or \textit{plural}), and these values can be either explicitly given (\textit{known}) or the target of a discovery action (\textit{unknown}). This model allows for more flexibility, providing a clearer picture of the task design space, and helps to estimate a task's overall complexity level considering the values of each dimension.

\subsection{Previous Evaluations of \highlight{Desktop-based} STCs}
\label{sec:rel:previous}

To date, most empirical evaluations of the STC compare it against equivalent 2D interactive maps and animations, investigating possible performance advantages or disadvantages in typical tasks. 
One of the first comparisons was presented by Kristensson et al. \cite{kristensson2009evaluation}, in a study with \highlight{30} novice users and four categories of tasks based on Andrienko et al.'s typology and a campus mobility use case. For simple tasks (e.g., ``Where is the red person at 2pm?''), they found lower error rates using the baseline 2D interactive map, while for more complex questions, which required comparison between multiple candidates (e.g., ``Who was on the campus for the longest time?''), the cube afforded similar accuracy but response times were twice as fast. 
Amini et al. \cite{amini2015impact} similarly compared an STC-based system, the Space-Time Visualizer, against a 2D interactive map, but employed different design choices and
tasks \highlight{in a study with 12 participants.} No differences in error rates were found, but their results indicate that the STC is beneficial in terms of completion time for tasks such as identifying meetings and stationary moments. They also analyzed the time spent on different actions, noting that participants spent considerable time rotating the STC, possibly to benefit from \textit{structure-from-motion} depth cues. 

In a more domain-specific study, Willems et al. \cite{willems2011evaluation} compared the STC to animations and density representations, investigating the visibility of vessel movement features under varying clutter. Their user study \highlight{with 17 participants} indicated that, for this domain and the chosen tasks, the density representation performed best, being outperformed by the STC only in determining the traffic lane with the most vessels. 

Kveladze et al. \cite{kveladze2013methodological} presented a series of usability evaluations of an extended analytical environment including the STC and other coordinated representations, employing multiple case studies and an iterative framework based on a series of phases, from the identification of system requirements by interviewing multiple domain experts (between 1 and 8 depending on the case study), to the proposal of design guidelines and conduction of expert 
user evaluations. They found that the STC was the most used representation in all tasks \cite{kveladze2015space,kveladze2017researching}. Experts \update{familiar with the data} reportedly performed better, probably due to their higher motivation to explore, 
but still complained about the steep learning curve and visual clutter of the STC---both issues that could potentially be minimized with a more intuitive immersive approach. 
As part of this study, Kveladze et al. \cite{kveladze2012we} also investigated the effectiveness of different visual variables, concluding that color schemes are useful to group trajectories, while the addition of depth cues based on shading creates too much clutter and should be avoided. 
\update{However, when evaluating data of different complexities \cite{kveladze2018cartographic}, color in combination with shading was the best performing option.} 

Gonçalves et al. \cite{gonccalves2016not} advocated the combined usage of the STC and 2D maps, 
since the analysis of trajectories is not limited to a single type of task. In user studies \highlight{involving between 16 and 30 participants,} they found that the 2D map is always faster and more accurate for location tasks, while the STC performs best for association tasks \cite{gonccalves2015cartographic}.

\begin{figure*}[t]
\centering
\includegraphics[width=.242\linewidth]{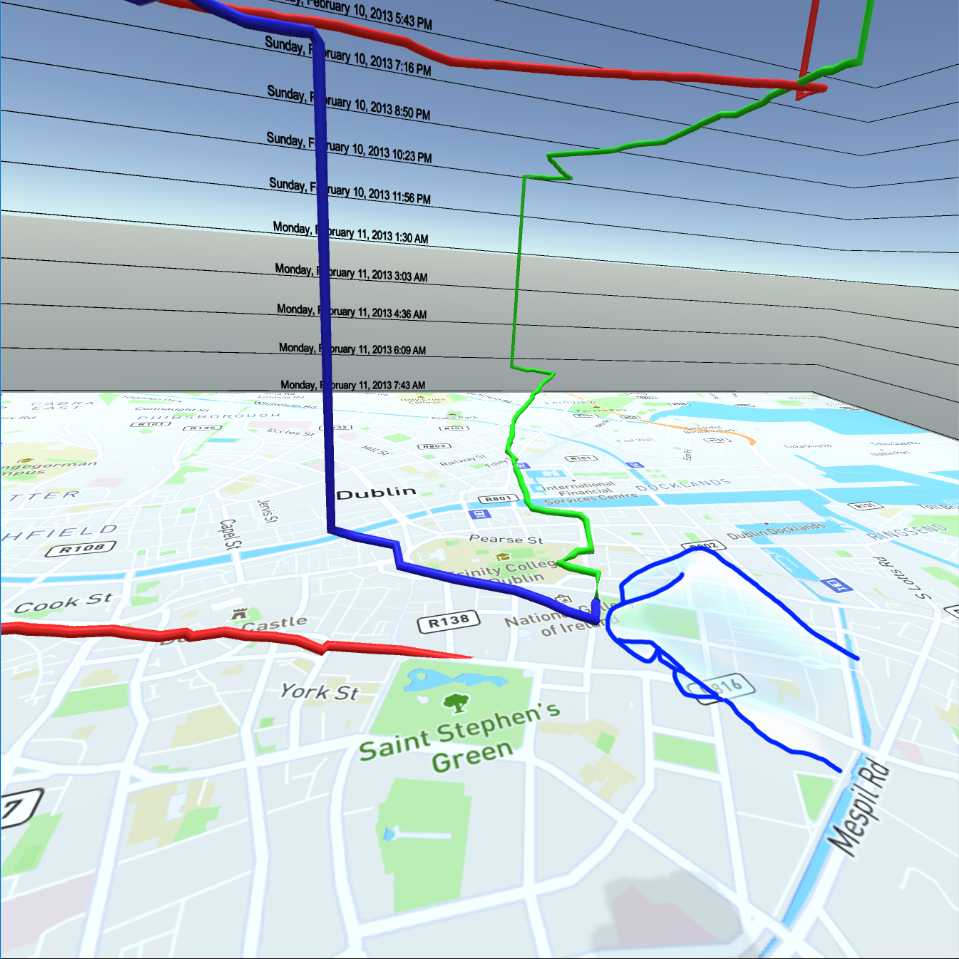}  \hfill
\includegraphics[width=.242\linewidth]{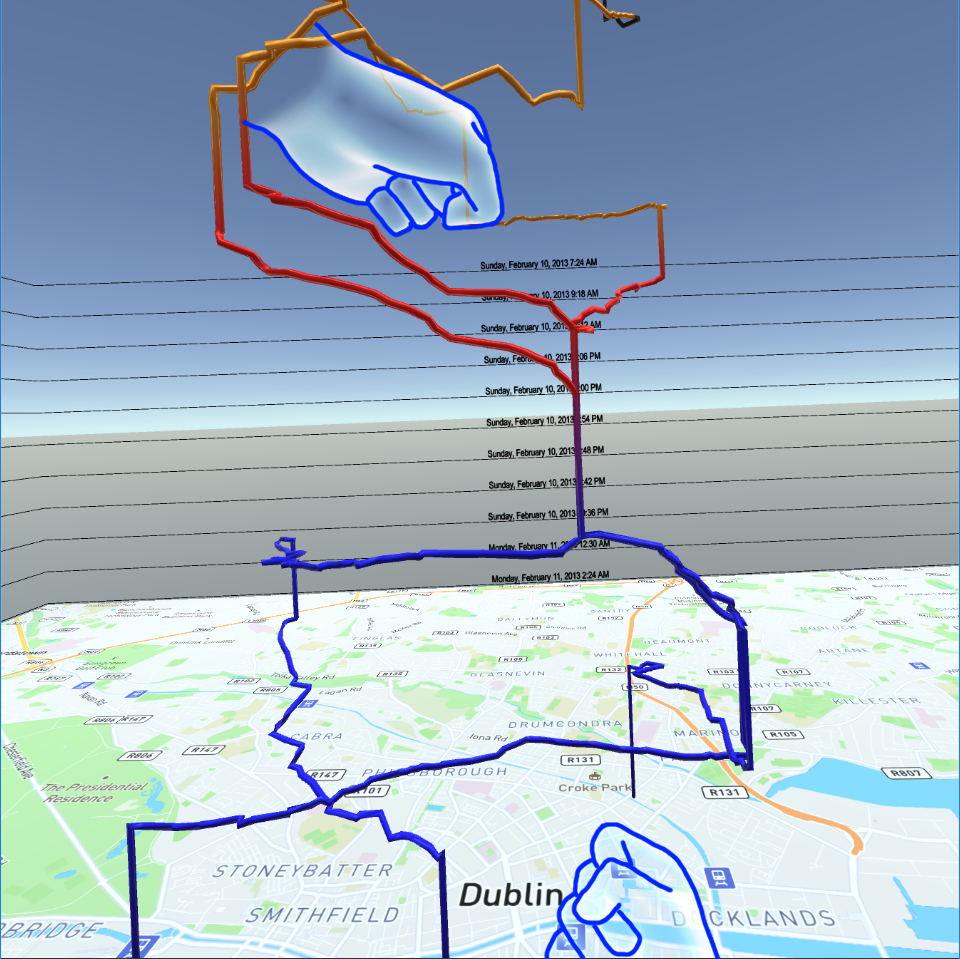}  \hfill
\includegraphics[width=.242\linewidth]{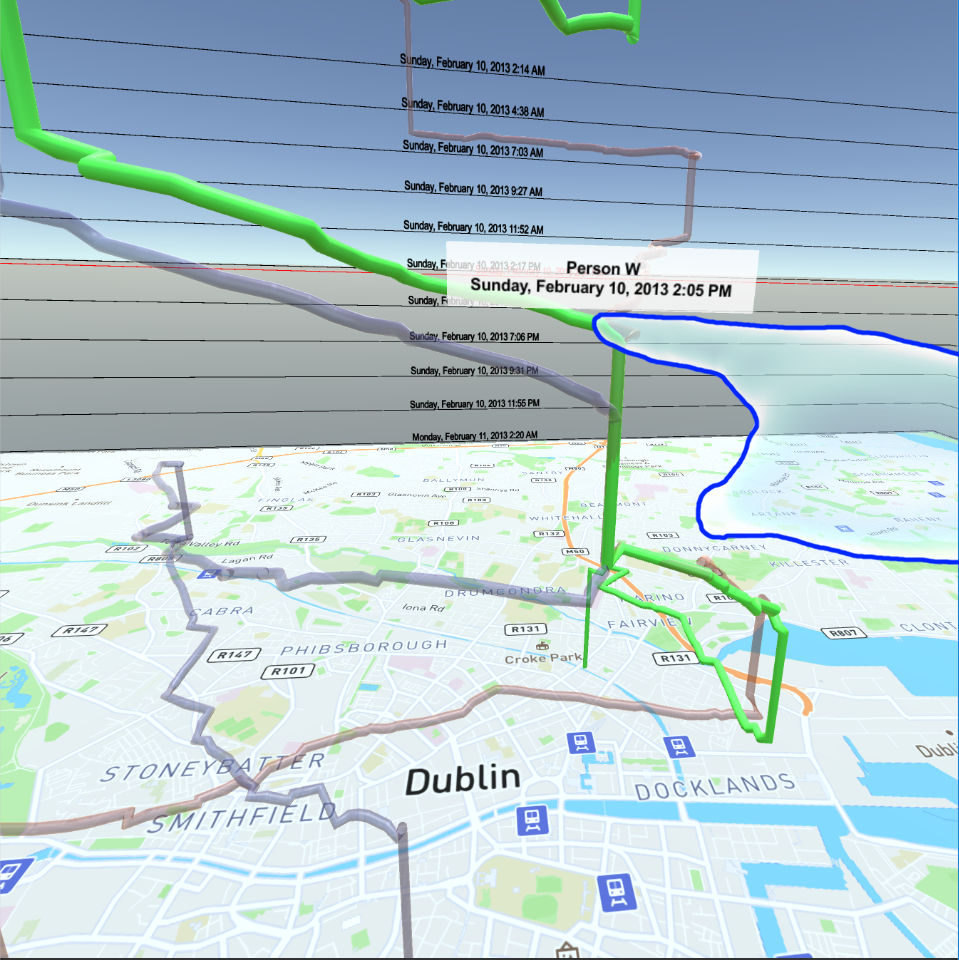}  \hfill
\includegraphics[width=.242\linewidth]{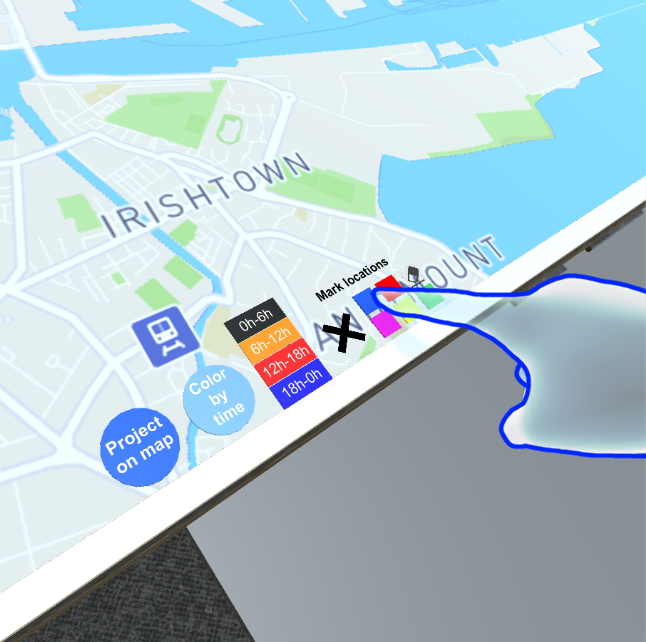}
\caption{In our immersive Space-Time Cube environment, all actions are implemented through intuitive mid-air gestures, such as grabbing (left), stretching (center left), and tapping (center right), or by tangible 
interaction with controls on the desk's surface (right). \highlight{Since time advances downwards, the trajectories show the movement history up to the point currently crossing the map. Blue hand contours added for clarity.}}
\label{fig:immersive}
\end{figure*}

\subsection{Immersive Analytics and the STC}

The hypothesis that stereoscopic displays could improve data exploration in an STC has been previously suggested, e.g., by Amini et al. \cite{amini2015impact}. Others, such as Kveladze et al. \cite{kveladze2013methodological}, have also emphasized the importance of depth (3D) perception for the STC. Nevertheless, little past work has actually explored this question.

Kjellin et al. \cite{kjellin2010different} compared monoscopic and stereoscopic versions of the STC, using a retro-projected table and polarized glasses in a disease spread event dataset \textcolor{black}{(i.e., with data points, but not trajectories).} Their results \highlight{from a study with 32 participants} indicated that stereopsis did not improve performance, and they claimed that the classes of structures that users need to perceive determine the applicability of the cube more than the depth cues used. Their tasks, however, consisted of identifying which data subsets spread in concentric circles (as real diseases), and which one had spread the most (i.e., had the largest diameter at the top level), making it possible to easily solve the task
via a top view of the cube with time encoded by point size. This means that they did not evaluate
the more complex data exploration tasks typical for STC use. Moreover, no qualitative evaluation was conducted. 
In subsequent work, the same stereoscopic STC was compared to a 2D map and animation for the visualization of trajectory data \cite{kjellin2010evaluating}. The STC was more precise and much faster for the determination of the order of arrival of different people at a meeting, while the map was more accurate for the prediction of a future meeting point of two trajectories. Yet, in this study the authors did not investigate a monoscopic version.

More recently, others employed systems with Head-Mounted Displays (HMDs) for STC visualization. Theuns \cite{theuns2017visualising} reported on focus group sessions discussing immersive prototypes of several movement data representations, including the STC. An unconventional ``map on the wall'' metaphor was used, with time depicted on a horizontal axis, and the users physically walked around to explore. Okada et al. \cite{okada2018} presented a large-scaled STC of Disneyland in VR, where geo-referenced tweets were grouped according to time and location, and the user inspected it by flying around.
Saenz et al. \cite{saenz2017reexamining}  reported on work-in-progress to implement an Augmented Reality (AR) STC for event data using the Hololens HMD. 
\update{Walsh et al. \cite{walsh2016braille,walsh2018tangible} discussed the potential of using purpose-built tangible controllers in Immersive Analytics for entity colocation in geospatial data. The STC was used in their system, but was not the focus of their study.} 
\highlight{Others, such as Nguyen et al. \cite{nguyen2017bees}, have introduced immersive representations of spatial trajectories, but without depicting temporal information.}
\highlight{None of these studies included evaluations of the STC with user studies, shedding little light into the specific questions investigated in our work.}

\highlight{\textit{GeoGate} is the work most similar to ours \cite{ssin2019geogate}. It is a system that combines a 2D tabletop display with a 3D AR ``hologram'' to visualize trajectories in the maritime domain. Due to the limited field of view of current AR HMDs, the STC is shown at a small size and its position is controlled by moving a tangible device, which works as a spatial filter. In a study with 24 participants, \textit{GeoGate}
reduced errors in tasks where users had to correlate different data sources. Due to its different focus domain and approach, 
that work is largely complementary to ours.}

These examples also illustrate the vast design space available for Immersive Analytics \cite{immersiveanalyticsbook}. Very different metaphors, such as virtual flying and physical walking, can be adopted, which can impact user experience and performance directly. 
One interesting option is to keep the analyst seated at their work desk \cite{zielasko2017remain}, using gestures to manipulate 3D data positioned above it. In a recent study, Wagner Filho et al. \cite{wagner2018virtualdesk} reported much lower completion times and simulator sickness incidence with a desk-based metaphor, called VirtualDesk, compared to flying. Taking into consideration the requirements of a domain analyst for a solution that is comfortable, efficient and easily integrable to their workspace, we \update{assume} here that a desk-based approach is more fit and adaptable for the immersive exploration of STCs.


\section{\highlight{Prototype} Immersive STC Environment} 
\label{sec:prop}

Here we introduce our approach for the immersive trajectory visualization environment that served as apparatus in our evaluation, as well as the rationale behind its main concepts and design choices for its implementation.

\subsection{\highlight{Interaction Design}}

Although other approaches, such as flying or room-scale walking, are possible and should also be investigated, we believe that a desk-based metaphor, such as VirtualDesk \cite{wagner2018virtualdesk}, is a natural fit for the STC. By allowing the user to remain seated and as no floor space is required, this approach is more easily integrable in current analyst's workspaces and workflows. Displaying the base map on the surface of their work desk is also a more natural metaphor and could aid the cognitive process, by mimicking a standard physical map exploration scenario (see \autoref{fig:teaser}). Moreover, \update{we selected a} one-to-one mapping between \update{gestural} interactions in the physical and virtual worlds, \update{instead of a laser-pointer-based approach,} to reduce the user's cognitive workload \cite{cordeil2017design}, and to afford higher precision, stronger stereopsis, and head-motion parallax through body-relative interactions 
within arm's reach \cite{mine1997moving}.

To allow a larger portion of the geographic space to be visualized,
we choose to make the virtual desk larger (3m\,\texttimes\,1.5m), taking advantage of the absence of space constraints in the virtual environment. This can also be seen as a focus+context approach \cite{cockburn2009review}, where the data 
shown on
the easily reachable part of the real desk is in focus and the remainder provides context.

\subsection{\highlight{Visualization Design}}
\label{sec:sys:design}

To \highlight{implement} an efficient immersive environment \highlight{for our user study,} we reviewed multiple interactive implementations of the STC (\autoref{sec:rel}), selecting the most appropriate design choices.

\highlight{\textbf{Time Direction.}} The first design choice concerns the direction in which time is displayed by default. While it may seem more natural in a 3D representation for the vertical axis to increase upwards, we agree with Amini et al.'s decision to display time increasing downwards \cite{amini2015impact}\highlight{, as also used in the \textit{GeoTime} system}~\cite{geotime2004}.  This way, the current center of inspection in time can be positioned to intersect the map's surface, while all previous movement that led to this moment remains visible above, instead of being occluded by the map (see \autoref{fig:immersive} -- left). 
Also, by moving the trajectories slowly upwards, the moment currently crossing the map advances and the user can visualize an intuitive animation of how the objects moved over time.
Grid walls were positioned around the desk \highlight{up to 1m of height} to complete the familiar STC environment and allow the analyst to quickly perceive 
(at least approximate) times for events, and also to indicate the current represented time granularity and direction. The user can also inspect any specific position with the aid of additional red horizontal lines, which permanently follow the movements of the hands when the index finger is raised (see \autoref{fig:teaser}).
To support other use cases, we offer an option to invert the vertical axis.

\highlight{\textbf{Moving the Trajectories vs. the Map.}} The second design choice concerns how different moments in time are inspected. Many implementations of the cube, including the original one by Kraak \cite{kraak2003space}, allow the transposition of a movable base plane across the trajectories to better identify object positions at a specific moment. Here, we chose to allow the user to move the trajectories themselves upwards and downwards. 
Compatible with our desk metaphor, this keeps the map 
on the desk surface, while 
also ensuring it is always visible and at a constant viewing angle \cite{amini2015impact}.  

\highlight{\textbf{Color Encoding of Time.}} Another choice was to allow the use of color encodings to represent the period of the day, as used by Gonçalves et al. \cite{gonccalves2015cartographic} (see \autoref{fig:immersive} -- center left). This feature is very helpful, for example, for the observation of recurring patterns and the quick identification of key locations, e.g., where each object spends their nights.

\highlight{\textbf{Trajectory Footprints.}} Finally, we also allow the projection of trajectory footprints onto the base map. This traditional feature of the STC \cite{kraak2003space} may be useful for tasks that do not require the temporal dimension, and also avoids 
the commonly observed strategy of users flattening the cube to obtain a 2D view \cite{gonccalves2015cartographic}.

\subsection{Mid-air and Tangible Interaction}

The exploration of the STC 
requires that the user is able to manipulate 
it to find 
an appropriate
view, and also to query its contents \cite{kraak2003space}. 
\highlight{With the goal of minimizing cognitive load by adopting a one-to-one mapping between real and virtual actions \cite{cordeil2017design},} interactions are based on two main gestures in our prototype: grabbing (with a closed fist, \highlight{see \autoref{fig:immersive} -- center left)} and tapping (with the index finger, \highlight{see \autoref{fig:immersive} -- center right)}. By grabbing with one hand, the user can translate the data in any direction, i.e., either through time or space. Moving two closed hands closer or apart scales the data either in time or space, altering the granularity at which that dimension is displayed. Spatial scaling is performed around the midpoint of the shown data, while time scaling is performed around the instant currently positioned at the map surface (assumed to be the current center of inspection). By rotating two hands, it is also possible to rotate the map to obtain different viewpoints. 
For inspecting trajectories, we use tapping actions (single tap), which shows a \textit{tooltip} panel with the trajectory's name and the exact timestamp (see \autoref{fig:immersive} -- center right) or select trajectories with a double tap, rendering the remaining ones semitransparent to facilitate analysis of specific data. Interactions with the trajectories produce haptic feedback on the controllers, to give better feedback for the mid-air content.

Our prototype affords a second category of interaction through controls displayed on the desk's surface, which makes it possible for users to tap them physically. 
These controls can be used to activate most of the features described in the previous subsection and also to select annotation colors from a palette (see \autoref{fig:immersive} -- right). Annotation colors become attached to the user's fingertip and act as ink, allowing pinpoints to be placed on any point of the map, or to fix wall lines at any instant in time by touching different parts of the trajectories. The purpose of annotations is to mark places or moments of interest during the exploratory process and these can persist for future analysis. 


\subsection{Technical Details}

The virtual environment was built in Unity3D and visualized through an Oculus Rift CV1 HMD (with two 1200x1080 displays). Oculus Touch hand controllers were used to track users' hands and to implement the hand gestures. 
However, all interactions are controller-agnostic, since they do not rely on any specific controller buttons 
\cite{wagner2018virtualdesk}. To minimize user discomfort \cite{yao2014oculus}, adequate computing and graphics hardware ensured a frame rate above 80 FPS. Dynamically zoomable maps were obtained from the Mapbox Unity SDK.

The trajectories were constructed as 3D tube meshes. Diameters were adjusted according to map zoom and started at 1cm. To optimize performance while preserving visual feedback, particularly in denser datasets, they were rendered with a lower level-of-detail, i.e., fewer vertices per position, during interactions, such as scaling, and then revert back to higher detail, when static for more than a second. 

Finally, we also updated the original VirtualDesk method of calibrating the desk's position by placing the hand controllers at a predefined location  \cite{wagner2018virtualdesk} 
with a new refined height adjustment method that asks the user to touch the desk's surface, which reduces the effect of any mis-calibration due to different real and virtual hand sizes.


\section{How to Evaluate an Immersive STC?}
\label{sec:howto}

Building on prior work that compared non-immersive STCs to other representations, we discuss here potential evaluation strategies.

\textbf{Measuring User Performance in Relevant Tasks.} 
Aided by relevant taxonomies (\autoref{sec:rel:questions}), the definition of representative tasks is essential for comparative evaluations, and should take the potential strengths of the compared systems into account. For comparisons between the STC and 2D interactive maps or animations, for example, selected tasks often require a good integration of space and time. 
In the case of the immersive STC, we believe that tasks with more requirement for interaction or involving comparisons between multiple objects and events are more likely to enable us to identify performance differences in comparison with a regular desktop-based STC, \highlight{due to the more intuitive, one-to-one mapping between real and virtual actions \cite{cordeil2017design}.} 
%

\textbf{Observing How Users Interact With It.} 
Amini et al. proposed examining the interaction process separately in terms of all features to gain an understanding of the causes behind the observed performance, and to isolate the \textit{static exploration} time in each condition \cite{amini2015impact}. 
Kveladze et al. also proposed a similar approach for the usage of different visual representations in their system, which they used to, e.g., deduce the participants' exploration strategies and to compare usage patterns of \update{expert users familiar and unfamiliar with the data \cite{kveladze2015space}.} 
%

\textbf{Interviewing Users About Their Subjective Experience.} Qualitative results obtained through standardized questionnaires and/or user interviews can be as important as the quantitative task results. Considering the previously reported steep learning curve for the STC \cite{kveladze2015space}, one of the hypothesized benefits of an immersive STC is better system usability. 
Besides easing the analyst's cognitive effort and allowing more attention to be paid to the actual information, a lower mental workload and higher usability can lead to higher adoptability of the STC. Yet, it is necessary to evaluate whether the lower familiarity of users with a VR environment does not result in a higher cognitive load.


%

\textbf{\update{Novice} Users vs. Domain Experts.} Both \update{novice} users and domain experts are important for evaluating a new visual representation or comparing a novel approach to an existing one. 
Domain experts are known to be more effective in the tasks due to their inherent motivation and familiarity with the data and analytical problems, and are able to perform more detailed analysis \cite{kveladze2015space}. For complex representations, such as the STC, they are also the main target users. 
Kveladze et al. recently proposed a methodological framework for assessing the usability of an STC which included several iterative steps integrating the domain experts before a final usability test
\cite{kveladze2013methodological}. \highlight{However, their focus was on understanding the utility of the STC for expert users in comparison to other representations.} 
\update{In our work, we build upon their result that STCs are beneficial for experts. Our goal is to improve user performance by increasing learnability, ease-of-interaction, and spatial understanding, which motivated us to conduct our evaluation with novice users.}
This choice allows us to assess a larger population 
and more accurately detect differences related to human perception and behavior. Moreover, it can also be argued that, in the words of Andrienko et al., nowadays \textit{``everyone is a spatio-temporal analyst''} in some way \cite{andrienko2010space}. 
Additionally, our tasks were designed to be atomic and easy to understand, without requiring specific domain knowledge.



\section{User Study}
\label{sec:study}

To investigate how our proposed immersive analytics environment performs in comparison with a conventional approach, we implemented a comparable desktop-based condition, and selected two data scenarios.
We also formally defined a set of tasks and applied the experiment design to a group of non-expert participants.

\begin{figure}[t]
\centering
\includegraphics[width=\linewidth]{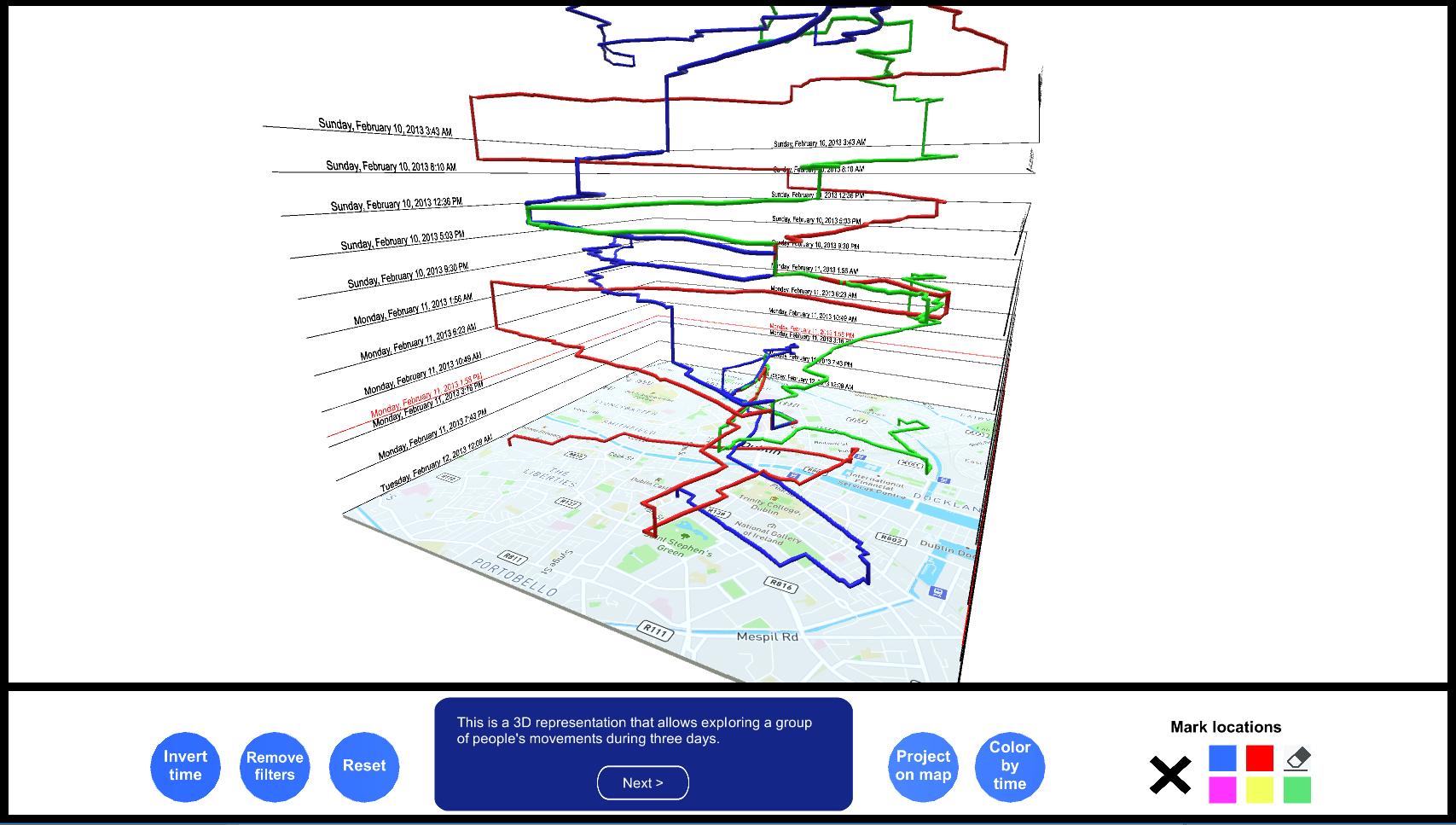}
\caption{The baseline condition builds on the same design choices
but with adaptations to fairly represent a typical desktop-based STC system.}

\label{fig:comparable} 
\end{figure}

\begin{table}[t]
\begin{center}
\begin{tabular}{C{.24\linewidth}|C{.31\linewidth}|C{.29\linewidth}}
\hline
\bf Action & \bf Immersive & \bf Desktop \\ \hline 
\bf Scale map & Move 2 closed hands & Scroll on map\\
\bf Scale time & Move 2 closed hands & Scroll on walls \\
\bf Translate map & Move 1 closed hand & Left-click drag map\\
\bf Translate time & Move 1 closed hand & Left-click drag wall \\
\bf Rotate map & Rot. 2 closed hands & Middle-click drag\\
\bf Rotate camera & Move head & Middle-click drag\\
\bf Translate cam. & Move head & Right-click drag, Ctrl+Scroll\\
\bf Inspect traj. & Index-finger tap & Mouse over \\ 
\bf Select traj. & Double tap & Double-click \\
\hline
\end{tabular}
\end{center}
\caption{While interaction commands in Immersive are mostly intuitive and body-related, Desktop mimics a typical Rotate-Pan-Dolly paradigm.}
\label{tab:interaction}
\end{table}

\subsection{\update{Hypotheses}}
\label{sec:study:hypos}

\update{We formulated 8 hypotheses for this study, concerning task performance (H1-H3), interactivity (H4-H5) and subjective experience (H6-H8).}

\begin{enumerate}
\setlength{\itemsep}{1pt}
\setlength{\parskip}{0pt}
\setlength{\parsep}{0pt}
\item[H1] The immersive STC will enable faster completion for tasks requiring much interaction.
\item[H2] The immersive STC will allow faster completion for tasks requiring the comparison between many possible alternatives.
\item[H3] In the immersive STC, stereopsis and proprioception will lead to better accuracy in difficult tasks.
\item[H4] In the immersive condition users will interact more frequently, due to the intuitivity of mid-air gestures.
\item[H5] In the immersive condition users will perform \highlight{fewer} rotations, as there is less need to perceive \textit{structure-from-motion}.
\item[H6] The immersive condition will lower the user's mental workload.
\item[H7] The immersive system will be rated as significantly more usable.
\item[H8] The desk-based immersive condition will be comfortable and not induce simulator sickness.
\end{enumerate}

\subsection{Baseline Condition}
\label{sec:study:comp}

The design for our baseline Desktop condition is based on the same 
choices, but uses standard mouse and keyboard controls for
interaction, see \autoref{tab:interaction} for the mapping. Some modifications were included to increase the compatibility with previous implementations of the STC in desktop settings, thus increasing the external validity. 
To this end, an extra degree of freedom for rotation was added, allowing the cube to be rotated up to 90$^{\circ}$ around its \textit{x} axis, enabling the user to obtain a birds-eye view. Such a rotation is not supported in the immersive version, since it would break the VirtualDesk metaphor. \highlight{Even though one could argue that this additional degree of freedom could increase the interaction effort, our observation of users indicates that this was actually an advantage for the Desktop system, as it supports user expectations for a 3D manipulation interface and even makes some tasks easier.} Note that we thus assume that the benefits introduced by the immersive metaphor compensate for any advantage resulting from this extra degree of freedom. 

The Desktop cube was also rendered in its more typical form with a square base, to keep the trajectories closer to a reference wall and \highlight{to fairly represent a typical desktop-based STC system.}
The STC manipulation was based on a typical 3D manipulation Rotate-Pan-Dolly paradigm \cite{jankowski2015advances}. 
A small strip at the bottom of the screen includes all commands that are available on the top of our virtual desk (see \autoref{fig:comparable}). This condition was experienced on a 22'' 1920x1080 display.

\begin{figure*}[t]
\centering
\includegraphics[width=.242\linewidth]{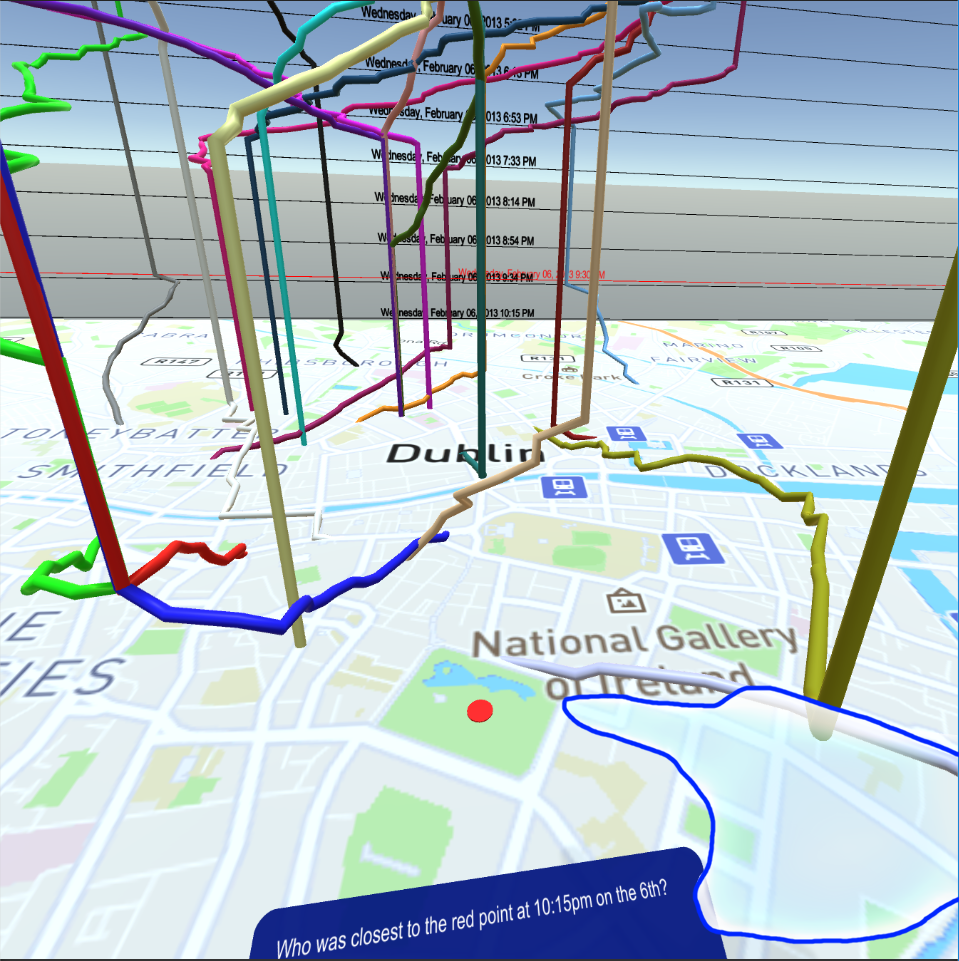}  \hfill
\includegraphics[width=.242\linewidth]{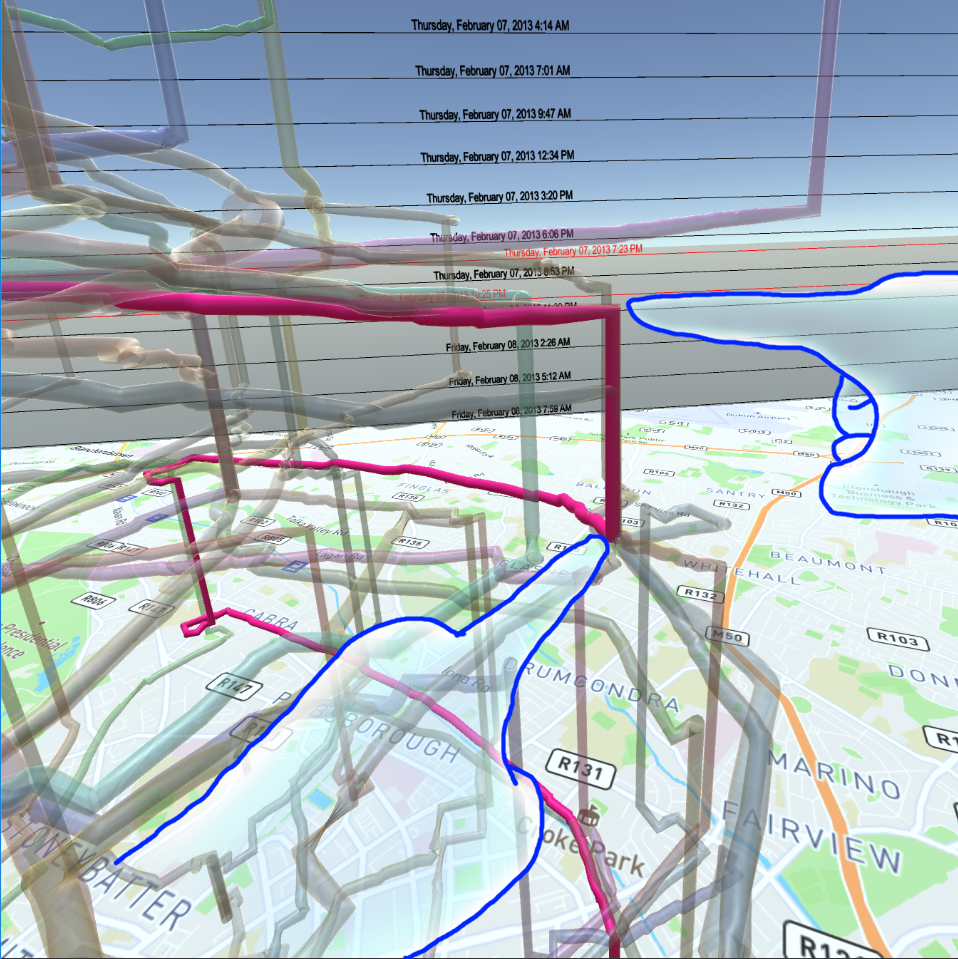}  \hfill
\includegraphics[width=.242\linewidth]{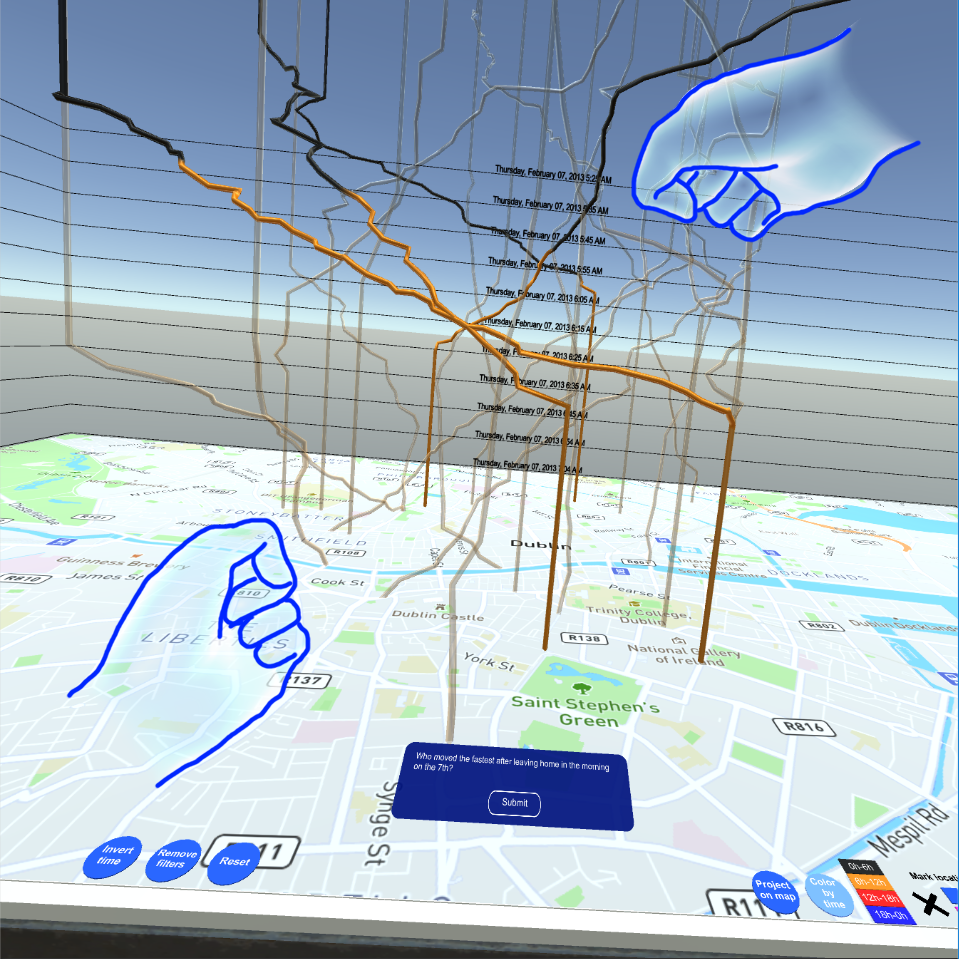}  \hfill
\includegraphics[width=.242\linewidth]{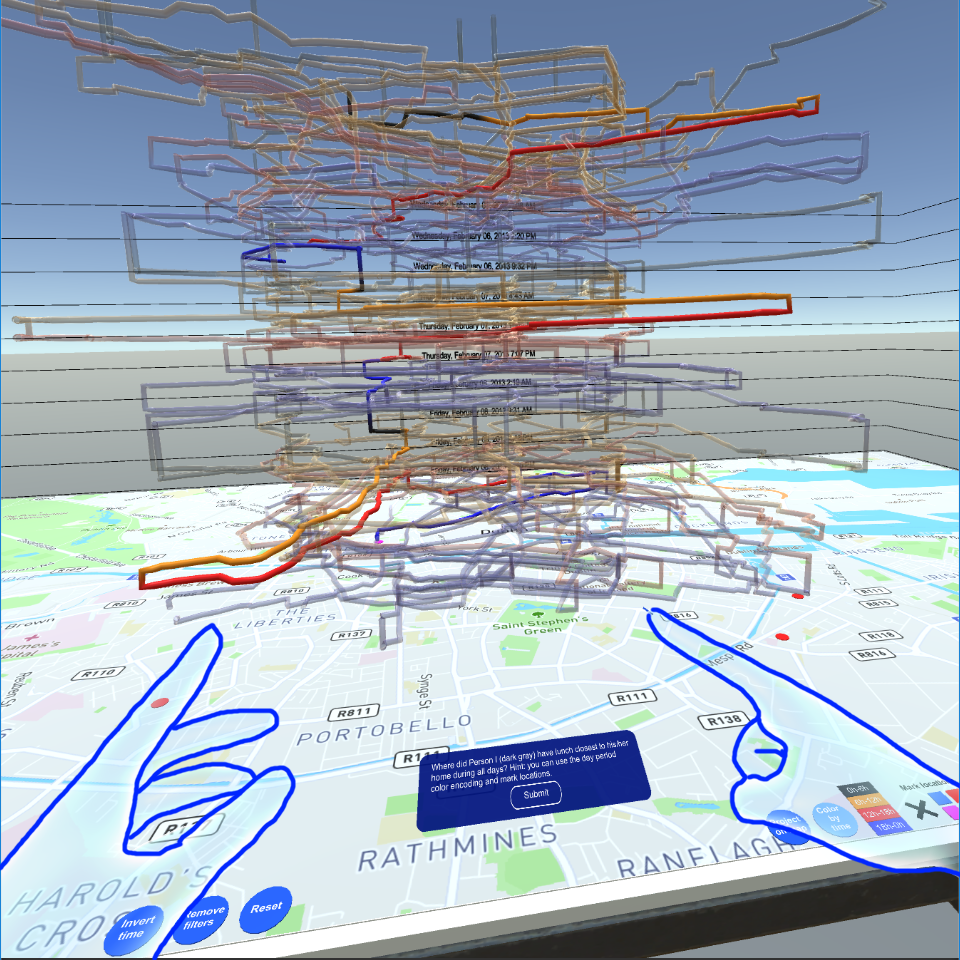}
\caption{Different tasks being performed in the Immersive condition 
with Dense data: comparisons of instant distance (left), stop durations (center left), movement speeds (center right) and event locations (right). \highlight{Blue hand contours added for clarity.}}
\label{fig:tasks}
\end{figure*}

\begin{table*}
\begin{center}
\begin{tabular}{c|C{2.6cm}|L{9.5cm}|C{1.0cm}|C{0.7cm}|C{0.8cm}}
\hline
& \bf Target Feature & \bf Example & \bf Object & \bf Time & \bf Space \\ \hline 
 \bf T1   & Instant position    & Where was Person A (red) at 10am on the 8th? & K,S & K,S & U,S \\ \hline
 \bf T2   & Meeting event, Time       & At what time did Person B (green) meet others for the first time on the 6th? & K,S & K,P & U,S \\ \hline
 \bf T3 & Stop duration    & Where did Person C (blue) spend less time on the 6th? & K,S & K,P & U,S \\ \hline
 \bf T4 & Events locations & Where did Person B (green) have his/her first lunch appointment closest to his/her home during all days? & K,S & U,P & U,S \\ \hline
 \bf T5 & Instant distance      & Who was closest to the red point on the map at 9am on the 8th? & U,S & K,S & K,S \\ \hline
 \bf T6 & Movement speed   &Who moved the fastest after leaving home in the morning on the 6th? & U,S & K,S & U,P \\ \hline
 \bf T7 & Order of events  & Who was the first to arrive at his/her lunch appointment on the 7th? & U,S & K,P & U,P \\ \hline
\end{tabular}
\end{center}
\caption{Evaluated tasks target different movement features, with different levels of difficulty and required interaction. Using Amini et al.'s approach \cite{amini2015impact}, the three movement task components are classified as \underline{K}nown or \underline{U}nknown, and \underline{S}ingular or \underline{P}lural.}
\label{tab:tasks}
\end{table*}


\subsection{Data Scenarios} 
\label{sec:study:data}

Due to their controlled nature and previous use in a similar study, we chose the simulated trajectory datasets by Amini et al.~\cite{amini2015impact} for our evaluation. 
These were created constructing a street intersections graph for the city of Dublin, Ireland, and obtaining shortest paths between predefined points. The 3-day dataset reproduces a controlled routine: each day, objects spend the night at their home points, leave for four different appointments in different locations during the day (morning, lunch, afternoon, and dinner) and then return back home. 

Taking into account that the presence of clutter is a deciding factor in the usability of the STC, with some reports that more than 10 trajectories result in excessive clutter \cite{gonccalves2015cartographic,demvsar2010space}, we decided to investigate two different data scenarios: a \emph{Simple} dataset with only 3 different trajectories (see \autoref{fig:immersive}), and a \emph{Dense} dataset with 24 simultaneous trajectories (see \autoref{fig:teaser} and \autoref{fig:tasks}). The latter was constructed by combining all datasets provided by Amini et al. Since the data spans a relatively small region of 10 square kilometers, the resulting set is very dense.

\subsection{Participants}
\label{sec:study:part}
A population of 20 graduate and undergraduate students (16 male/4 female, mean age 22.2, SD 3.2) was recruited from our University campus. Half the users presented corrected-to-normal vision and wore glasses in combination with the HMD. 
13 
reported no or low previous experience with VR HMDs, while only 5 were very experienced. Nonetheless, all participants reported at least average experience with 3D computer games, 17 with gamepads, and 12 with motion controllers, characterizing a population familiar to navigation in 3D environments.

\subsection{Evaluated Tasks} 
\label{sec:study:tasks}

We defined seven tasks that aim at different target movement features, levels of difficulty, and required interaction. They are presented and classified in \autoref{tab:tasks}. 
Tasks 1 to 4 concern a single trajectory, with relatively similar difficulties in both data scenarios (the target object can be selected, which renders the remaining ones semitransparent and minimizes the influence of clutter). Tasks 5 to 7 require the comparison of all objects, which becomes much more difficult in the dense scenario. 
We selected two different task stimuli points or periods for each task in each data scenario, targeting similar situations and with similar degrees of difficulty. In all cases, it was reasonably easy to perform all required selections, and tasks required a reasonable degree of comprehension and exploration to extract the information from the STC. All tasks had an objective known answer which was used to assess user choices without introducing subjective criteria. For tasks which required the selection of a time or location, we adopted an error margin of 15 minutes or 2 centimeters on the map's surface.

The procedure was similar for all tasks: users must first identify the target element or period, then move the desired time or period closer to the map plane and identify the answer or compare 
different possibilities. 
We designed Task 4 to force the exploration of multiple regions of the data. There, users were expected to find the home site of the target subject and each of the three locations where they had lunch, marking them on the map for a later comparison. Thus, before the start of this task, participants were explicitly reminded about the availability of map annotations and 
color encodings.


\begin{figure*}[t] 
\centering
\includegraphics[width=\linewidth]{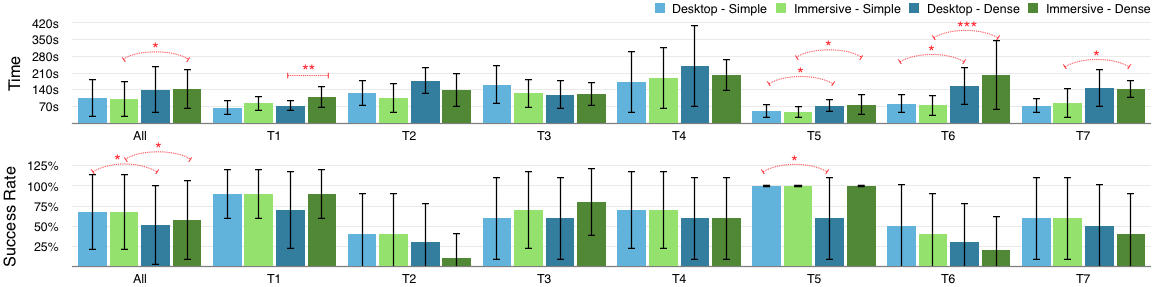}
\caption{Times were similar in both conditions, with the exception of the simplest task. Success rates were generally similar across scenarios and conditions, except for T5, which became more difficult for Dense data in Desktop, but not in Immersive. \highlight{Error bars indicate standard deviation.}}
\label{fig:res:microperf} 
\end{figure*}

\subsection{Experiment Design}
\label{sec:study:exp}

The experiment followed a mixed design, with 2 within-subjects \textit{visualization conditions}\,\texttimes\,2 between-subjects \textit{data scenarios}\,\texttimes\,7 \textit{tasks}, totaling 14 trials per participant with an average duration of 65 minutes including the familiarization phases and questionnaires filling. 
The need for a between-subjects independent variable became clear in pilot testing, where we observed that, even for a very small number of tasks, untrained users required a long time. Increasing the test length would introduce additional noise resulting from participant fatigue and also lead to a very long VR exposure period. Considering that the visualization condition is our main investigated variable, we prioritized it as within-subjects to avoid the interference of interpersonal differences. Participants were randomly assigned to one of the data scenarios at the beginning of their tests. 
The presentation of conditions was alternated to counterbalance learning biases. Task stimuli were also alternated between conditions after each test to balance occasional unintended difficulty differences. Task order was always preserved. 

At the beginning of each condition, users received a guided tour of the main system features, and also instruction for how to interpret the cube, since it is a representation unfamiliar to most users. During this period, which lasted on average 8 minutes for Immersive and 6.2 minutes for Desktop, they freely explored a practice dataset with three trajectories and were asked to practice finding simple information, such as an object's position at a given time. 
During the experiment, task questions were presented via text at the bottom of the screen in Desktop and on a small panel above the desk's surface in Immersive. Participants could read and ask for clarifications before pressing a button to start and the trajectories were only shown after this moment.
Participants were always asked to prioritize accuracy over 
speed in their answers. 
The supplementary videos demonstrate all system functionality and how tasks were performed in both conditions.

After experiencing each condition, participants were asked to fill standardized questionnaires, including the System Usability Scale (SUS) \cite{brooke1996sus}, the NASA Raw TLX for workload assessment \cite{hart2006nasa}, and the Simulator Sickness Questionnaire (SSQ) \cite{kennedy2003configural}, where the latter was applied pre and post VR exposure. 


\section{Results}
\label{sec:res}

We report here the user study results in terms of quantitative measures, qualitative performance, interactivity and participant feedback. 
\update{Since most variables were not normally distributed,} paired or unpaired \update{non-parametric} Wilcoxon signed-rank tests\update{\cite{wilcoxon1945individual}} were used and significance is indicated 
as follows: ($*$) for \textit{p} $<$ 0.05, ($**$) \textit{p} $<$ 0.01 and ($***$) \textit{p} $<$ 0.001. 
We also report \textit{Z}-values and effect sizes (\textit{r})~\cite{pallant2013spss}.



\begin{figure}
\centering
\includegraphics[width=\linewidth]{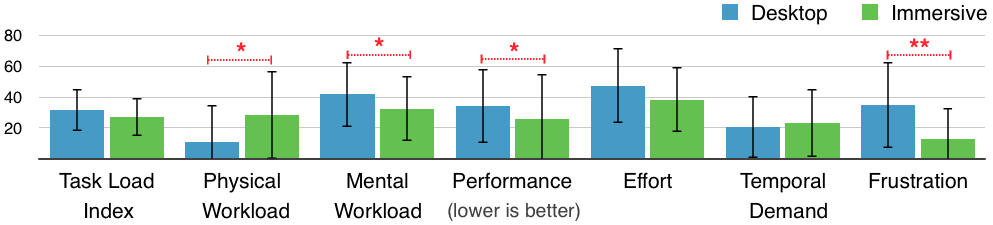}
\caption{Task workload components of the Nasa TLX questionnaire under Desktop and Immersive. Mental Workload and Frustration were significantly lower in the latter. \highlight{Error bars indicate standard deviation.}}
\label{fig:res:tlx} 
\end{figure}

\begin{figure*}[t] 
\centering
\includegraphics[width=\linewidth]{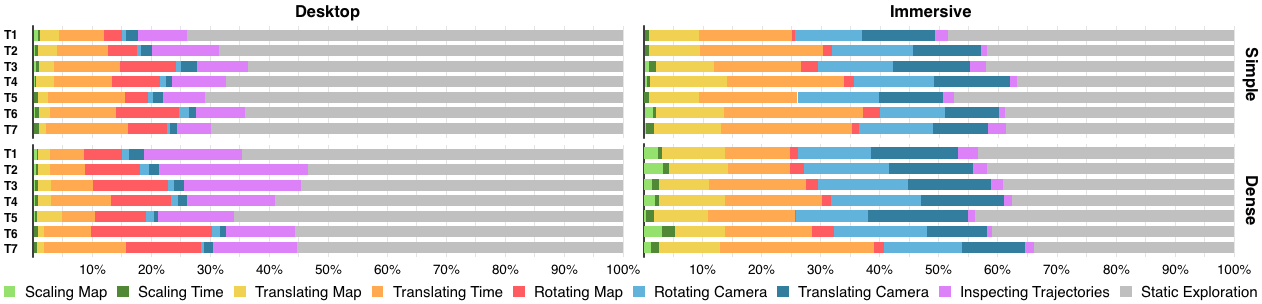}
\caption{Distribution of time across the different interaction features, for tasks in both conditions and scenarios. Users in Desktop performed many more map rotations, to benefit from \textit{structure-from-motion}, and inspections, to be able to accurately determine times. In Immersive, users intuitively performed much more data translations and scalings, and constantly moved their heads to change their point of view.}
\label{fig:res:interact} 
\end{figure*}

\subsection{Task Performance}
\label{sec:res:perf}

\autoref{fig:res:microperf} presents averages and standard deviations for elapsed times and success rates, overall and per task. 
Few significant differences were observed, contradicting hypotheses H1-H3. 

In terms of time, T1 (instant position) was slightly slower in Immersive for both Simple (\textit{p} = .08, \textit{Z} = -1.7, \textit{r} = .38) and Dense (\textit{p} = .005, \textit{Z} = -2.8, \textit{r} = .61) datasets. In Desktop, this very basic task could be answered by positioning the required time on the map's surface and selecting the position from a distance, while in Immersive the selection requires first reaching that location. On the other hand, T3 (stop comparison) was faster in Immersive for the Simple dataset (\textit{p} = .08, \textit{Z} = -1.7, \textit{r} = .38).

In terms of success rates, we found only a single significant difference. For T5 (instant distance), increasing data density in Desktop significantly decreased the correct rate (\textit{p} = .03, \textit{Z} = -2.1, \textit{r} = .47), while for Immersive no differences were found. 
The low success rate presented by Immersive in task T2 for Dense data was partially contaminated by unintended selections, given that time annotations were performed by a single tap and trajectories were very close in this scenario. Still, there was no significant difference, with \textit{p} = .4.

As expected, tasks T5-T7, which require comparisons between all objects, presented a significant slowdown in the Dense condition. Only T5, however, presented fewer successes, and only for Desktop (\textit{p} = .03, \textit{Z} = -2.1, \textit{r} = .47).

\subsection{Usability, Comfort and Workload}
\label{sec:res:subj}

Desktop and Immersive presented significantly different usability evaluations on the System Usability Scale (SUS). While the former was rated with an average score of 62.1 (SD 20.6), the latter received 82.3 (SD 10.1), confirming our hypothesis H7 ($***$).

In terms of workload as measured by the NASA Raw TLX questionnaire, results were also favorable to the immersive system (see \autoref{fig:res:tlx}), with a non-significant smaller overall index (26.8 vs. 31.6, \textit{p} = .07) but significantly lower Mental Workload (32.4 vs. 41.6, $*$), which supports H6. Frustration was also greatly reduced (from 35 to 12.5, $**$) with immersive interaction, while the Physical Demand naturally increased (from 10.8 to 28.3, $*$). Moreover, perceived performance was also significantly better in Immersive (25.8 vs. 34.1, $*$, where lower is better). Comparing participants 
across both data scenarios, the only 
significant difference is the Mental Demand for Desktop, which was 31.6 in Simple and 51.6 in Dense ($*$), while participants in Immersive did not present a significant difference. 

Finally, no significant difference was found between pre and post-VR Simulator Sickness scores (\textit{p} = .22), with an average delta of 2.8 (SD 16.5), considered negligible \cite{kennedy2003configural}, confirming H8. Furthermore, no user reported discomfort, despite an average of 25
minutes of VR exposure (SD 8.2, \textit{max} = 45.16 min).

\subsection{Interactivity Analysis}
\label{sec:res:inter}

\autoref{fig:res:interact} illustrates the distribution of users' times across all possible interactions. For this analysis, camera rotations and translations in the Immersive condition refer to head movements (above a small threshold \update{empirically determined as 1$^{\circ}$ and 1~mm, respectively),} and, in Desktop, to vertical rotations around the cube and strafing/zooming. Camera rotations around the vertical axis in Desktop are seen as map rotations. When no specific interaction was detected, we assumed that users were performing \textit{static exploration} of the data \cite{amini2015impact}.

In this data, differences in terms of multiple interactions are evident and participants appear to perform many more interactions in the Immersive condition (H4). Camera movements are much more common in Immersive, where users spend a large part of their times rotating and moving their heads to look at different points and from alternative points of view. While zooming, strafing, and vertically rotating the cube are also important features in Desktop, a very small fraction of interaction time is dedicated to these actions. On the other hand, map rotations (which in Desktop correspond to actually rotating the whole cube around its vertical axis) are one of the most performed interactions in Desktop, confirming previous observations from Amini et al. \cite{amini2015impact}, while they are rare in Immersive (confirming H5). Such rotations were even more frequent in Desktop in the Dense scenario, possibly to obtain different points of view and to avoid occlusions. 

Another notable difference is the much higher number of trajectory inspections in Desktop, which display a \textit{tooltip} with the exact time of any point. Since this feature is activated by placing the mouse over any trajectory point, this can be performed from a distance at any time (sometimes even unintentionally). In contrast, in Immersive the user would have to tap at the 3D location. This can also be partially attributed to the much greater difficulty of identifying the correct times of points using the cube walls in Desktop than in Immersive, where users have an egocentric point of view. 


\subsection{Participant Feedback} 
\label{sec:res:feed}

\begin{figure}
\centering
\includegraphics[width=\linewidth]{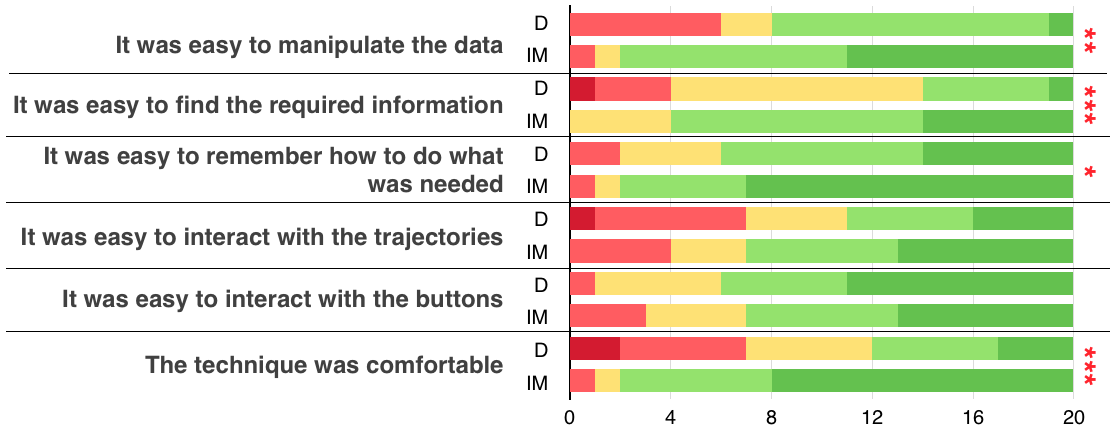}
\caption{Likert-scale agreements to different assertions \highlight{ranging from strongly disagree (dark red) to strongly agree (dark green).} Participants found it significantly easier to find information and interact in Immersive \highlight{(IM) than in Desktop (D),} and also considered it more comfortable.}
\label{fig:res:likert} 
\end{figure}

\highlight{Participants were also asked to provide feedback through Likert-scale questions (see \autoref{fig:res:likert}) and open comments. 
We observed that for four out of six assertions presented, agreement rates were significantly larger for Immersive. 
For Immersive, 90\% of participants agreed that it was easy to manipulate the data and to remember how to do what was needed, 
80\% 
that it was easy to find the required information, 
and 90\% that the technique was comfortable, 
in line with the usability results (\autoref{sec:res:subj}). 
The only questions which did not see significance were related to interactions with the buttons and trajectories. Yet, considering that physically 
\update{touching} a button on the desk or selecting a trajectory mid-air could be potentially more difficult, this is also positive and the results for these questions are probably related to challenges with selections and in understanding the meanings assigned to the buttons. 
}

In open comments, several participants reported that the immersive environment was ``more fun'' and ``more interesting''. 
However, one participant mentioned that it was also ``tiring'' to scale and rotate the data, and some complained that, using gestures rather than a mouse, it was more difficult to perform precise actions, such as moving a specific time to the desk surface. 
One explicitly justified that this could be due to their little experience with VR, but this can be \update{seen} also a limitation of using mid-air manipulation gestures. With respect to Desktop, one participant mentioned its controls were similar to the game TheSims, which made it easier to use. 

A feature that caused difficulty for many users was the double-tapping gesture. To avoid unintended selections and confusion with the single tap inspection, our implementation of double tapping required precision in tapping frequency and location, which mimics a mouse double-click. However, perfecting this mid-air action requires practice, and, in contrast to our expectations, users often became less and less precise with time, due to their frustration. Yet, no significant difference against Desktop was found in terms of ease to interact with trajectories (\autoref{fig:res:likert}). 

Some users had difficulty with rotations, which required both hands to be at the same height to avoid confusion with temporal scaling, and complained about this constraint. 
They also often tried to translate the data by moving two closed hands, which was again constrained to avoid confusion with spatial scaling and rotation. However, when multiple actions were assigned to the same gesture, such as spatial and temporal translation with one closed hand, some also \highlight{paradoxically} complained about the difficulty to perform one without affecting the other.

Finally, participants were asked to rank the two conditions on different criteria. 19 judged Immersive to be more engaging, 18 to be more intuitive, and 13 also deemed it the fastest. In terms of being the most accurate, half the participants chose each condition, which is possibly linked to the difficulty of executing precise movements in Immersive. 

\section{Discussion}
\label{sec:disc}

\highlight{Here we discuss our findings to extract recommendations for future work in the STC community, and possible explanations for why some of our initial hypotheses could not be validated at this time.}

\subsection{Findings for the STC Community}

In comparison to the conventional desktop-based approach, our results indicate several clear qualitative benefits for the exploration of trajectories with immersive STCs. 
The 32\% larger usability score and 32\% smaller Mental Workload index indicate that immersion could be a significant improvement for the acceptability of the STC, which partially addresses the reported steep learning curve and cumbersome manipulation. This was confirmed by the Likert-scale data, where 16/20 considered it easy to find the required information and 18/20 considered this technique comfortable, compared with 6/20 and 8/20 in Desktop, respectively. \highlight{Even though this is not the first time an immersive visualization system has presented lower workload than its
desktop-based counterpart \cite{wagner2018virtualdesk}, each specific representation has its own particularities in terms of required interactions and information-seeking activities, demanding an independent evaluation. 
Considering the large number of possible interactions and low previous familiarity of participants with the proposed interactions, and even though we expected to see such behavior (H6 and H7), we were still surprised with the large observed 
differences between conditions.} 
Moreover, the immersive system exhibited similar Mental Workload and success rates for all tasks in the Simple and Dense data scenarios, in contrast to Desktop, which was more susceptible to the presence of clutter. \highlight{This observation could guide STC users when selecting the best approach for their analysis depending on data complexity.} 

\highlight{In terms of visualization design, our observations indicate that our choices 
were successful in providing an easily understandable STC implementation, 
even for novice users, and we recommend them for future similar studies. The downwards direction of time, albeit initially confusing, was generally appreciated 
due to the possibility of following an object's position on the map by dragging the trajectories upwards. The color encoding of time was very useful for task T7, which required the identification and comparison of lunch locations and times.}

\highlight{For our interaction design, user feedback demonstrates that our choices were successful in enabling easier and more intuitive exploration of the data (see \autoref{sec:res:feed}), but also indicates directions for future improvements. First, interaction gestures cannot suffer from (seemingly) unnecessary constraints or ambiguities that contradict users' expectations.
Visual or haptic feedback could be used to deal with such situations. Second, the requirement for direct selection of points by double tapping resulted in small time overheads in task T1. Remote point selection functionality without breaking the interaction metaphor is a topic for future work. Finally, we could consider alternative mid-air manipulation techniques for map zoom-and-pan to minimize potential arm fatigue during longer use, as discussed by Satriadi et al. \cite{satriadi2019freehand}.}

The intuitive interaction scheme deeply affected the interaction patterns. 
Besides constantly moving their heads to obtain different points of view, users performed more data translations and scaling using gestures. Yet, the total time was still similar to Desktop (with the exception of T1, the simplest task, which was slightly faster in Desktop). In Desktop, users based their exploration on frequent map rotations to better perceive the scene. 
Differences in data translation over time or space can probably be attributed in a large part to the intuitivity of the grabbing action, which stimulated its frequent usage, but also to the need to first move the data and then select, inspect, or annotate it, which is not required in Desktop. 
Finally, data scaling was also noticeably more common in Immersive and in the Dense scenario. Scaling is an important feature to increase the spacing between objects in a dense representation. The fact that it was not more frequently used in Desktop, despite mouse scrolling being a standard command for zooming, is surprising. \highlight{Yet, based on our observations and the requirements of our specific tasks, we have no reason to believe that the reduced resolution in the immersive condition is a possible explanation for the results.}
\highlight{Instead, we observed that users felt more inclined to use scaling to ``dive'' into the data in Immersive, when they deemed it necessary.}

Keeping in mind that our 
participants are accustomed to 3D navigation, we expect that difference for other user groups, such as data specialists, could potentially be even larger. 
Furthermore, results for simulator sickness were also excellent, despite relatively long sessions. This concern seems to be a non-issue for our desk-based VR metaphor.

\subsection{Why Was Quantitative Performance Not Improved?}

Contrary to our initial hypotheses H1-H3, few differences were found in user performance. We believe 
several factors contributed to this. First of all, participants were unfamiliar with the STC. According to our observation, error rates increased when difficulties to correctly interpret the representation arose. Also, times increased due to difficulties in understanding how to efficiently extract the required information. Moreover, despite our intention to explore a wide range of task complexities, the conceptual difficulty experienced by many users led to tasks being always either too easy or too difficult, which becomes clear when looking at the success rates \autoref{fig:res:microperf}. Yet, providing more training would render test sessions very long, introducing fatigue noise. 
Finally, and even though two of our hypotheses refer to time, we asked users to prioritize being accurate to avoid rushed or careless answers. This led many to lose focus during the exploration, especially in the immersive condition, which was more novel and interesting to them.

Our participants were previously familiar with 3D navigation in computer games and generally not familiar with VR environments, but still preferred the latter and considered it easier. 
Moreover, it is positive that Immersive did not increase times despite adding an additional level of new information to be learned. In summary, considering the population of our study and the complexity of the tasks evaluated, we are not completely surprised that we could not observe significant improvements in terms of quantitative performance.

\subsection{Limitations}

A main limitation of our study was the fact that a \update{novice} population was assessed, with no previous familiarity or training with the STC, which constrained the observation of quantitative performance. Participants required an average of 65 minutes to complete only 14 trials, 
which can be considered very long for this kind of experiment. 

In terms of the technique, one limitation detected by participants was the difficulty to apply precise adjustments, for example in data translation, as well as to translate time and space independently, actions which are better supported by Desktop. On the other hand, when gestures were constrained to increase precision, 
users were also not satisfied. This indicates that gestures should be made as natural as possible to correspond to users' expectations, but more specific techniques should be considered for cases when precision is required. 
One possibility is the use of dedicated buttons on the desk's surface to enable 
adjustments with fixed increments, which could also reduce physical demand.

Further, several participants experienced difficulties with the double-tapping action. To avoid confusion with simple inspections (which correspond to a mouse-over), our implementation necessitated high precision in frequency and position of the taps. 
This action must be revisited to enhance usability and lower training requirements, while still preserving precision. However, we are confident that the gesture is appropriate for selection, maintaining the desired embodied metaphor, in line with the remaining interactions in the prototype. One evidence for this is that 
participants still rated the interaction with trajectories better in the immersive environment. The importance of double-tapping also became evident when considering the high number of unintended annotations 
(requiring only a single-tap action), which partially contaminated results for task T2 (\autoref{sec:res:perf}).

\update{Finally, the results 
are limited to the design choices made in our study, and alternative approaches, e.g., an egocentric room-scale environment, should be assessed in future work.}

\subsection{Perspectives for Future Work}

We see many perspectives for the immersive STC. Besides an evaluation with domain experts, one is the expansion to other categories of spatio-temporal data, such as event and origin-destination datasets, both of which will introduce different requirements, and benefit from our approach in different ways. 
Further evaluation employing alternative datasets, including real-world data, are also planned. In particular, we
are considering
the exploration of big trajectory datasets for collective behavior exploration 
\cite{kveladze2012recognition,kveladze2015space}. 
To this end, novel trajectory mesh rendering techniques are needed to preserve the essential high interactivity and frame rate required for VR comfort \cite{hurter2018fiberclay}. 

Further, we also plan to expand our immersive environment to include complementary data representations and support more task categories \cite{kveladze2017researching, gonccalves2016not}. We might also investigate additional information encodings, such as varying tube radii, 
and the integration of desktop-based and immersive implementations, to allow seamless transitions, according to the analyst's needs \cite{wagnerfilho2019cga}.

\section{Conclusion}
\label{sec:conc}

In this work, we evaluated an immersive implementation of the STC geovisual representation based on the VirtualDesk exploration metaphor, aiming to increase its usability and to decrease its learning curve. To this end, we reviewed previous literature in Immersive Analytics and Geovisualization and made a series of design choices. 

Even though our work can only be considered a first step into the investigation on the benefits of immersive STCs, we were able to report very promising results for the qualitative experience, with much higher usability and user preferences, very low simulator sickness incidence, and a lower measured mental workload. The next steps for our research include further detailed evaluations with domain experts and the application of our approach to real-world spatio-temporal datasets.


\acknowledgments{
The authors thank the user study participants for their feedback and the reviewers for their insightful contributions. We also acknowledge financial support from CNPq--Brazil and Global Affairs Canada. This study was financed in part by the Coordenação de Aperfeiçoamento de Pessoal de Nível Superior - Brasil (CAPES) - Finance Code 001.} 


\bibliographystyle{abbrv-doi}

\bibliography{template}

\begin{thebibliography}{10}

\bibitem{amini2015impact}
F.~Amini, S.~Rufiange, Z.~Hossain, Q.~Ventura, P.~Irani, and M.~J. McGuffin.
\newblock The impact of interactivity on comprehending {2D} and {3D}
  visualizations of movement data.
\newblock {\em IEEE Transactions on Visualization and Computer Graphics},
  21(1):122--135, 2015. doi: {{%
10\hspace{.1pt}\discretionary{.}{%
}{.}\hspace{.4pt}1109\discretionary{/}{%
}{/}TVCG\hspace{.1pt}\discretionary{.}{%
}{.}\hspace{.4pt}2014\hspace{.1pt}\discretionary{.}{%
}{.}\hspace{.4pt}2329308}}


\bibitem{andrienko2010space}
G.~Andrienko, N.~Andrienko, U.~Demsar, D.~Dransch, J.~Dykes, S.~I. Fabrikant,
  M.~Jern, M.-J. Kraak, H.~Schumann, and C.~Tominski.
\newblock Space, time and visual analytics.
\newblock {\em International Journal of Geographical Information Science},
  24(10):1577--1600, 2010. doi: {{%
10\hspace{.1pt}\discretionary{.}{%
}{.}\hspace{.4pt}1080\discretionary{/}{%
}{/}13658816\hspace{.1pt}\discretionary{.}{%
}{.}\hspace{.4pt}2010\hspace{.1pt}\discretionary{.}{%
}{.}\hspace{.4pt}508043}}


\bibitem{andrienko2003exploratory}
N.~Andrienko, G.~Andrienko, and P.~Gatalsky.
\newblock Exploratory spatio-temporal visualization: An analytical review.
\newblock {\em Journal of Visual Languages \& Computing}, 14(6):503--541, 2003.
  doi: {{%
10\hspace{.1pt}\discretionary{.}{%
}{.}\hspace{.4pt}1016\discretionary{/}{%
}{/}S1045\discretionary{%
}{-}{-}926X\discretionary{%
}{(}{(}03\discretionary{)}{%
}{)}00046\discretionary{%
}{-}{-}6}}


\bibitem{bach2014review}
B.~Bach, P.~Dragicevic, D.~Archambault, C.~Hurter, and S.~Carpendale.
\newblock A review of temporal data visualizations based on space-time cube
  operations.
\newblock In R.~Borgo, R.~Maciejewski, and I.~Viola, eds., {\em Eurographics
  Conference on Visualization - STARs}. The Eurographics Association, 2014.
  doi: {{%
10\hspace{.1pt}\discretionary{.}{%
}{.}\hspace{.4pt}2312\discretionary{/}{%
}{/}eurovisstar\hspace{.1pt}\discretionary{.}{%
}{.}\hspace{.4pt}20141171}}


\bibitem{bertin1983semiology}
J.~Bertin.
\newblock {\em Semiology of Graphics: Diagrams, Networks, Maps}.
\newblock ESRI Press, 2011.

\bibitem{brooke1996sus}
J.~Brooke.
\newblock {SUS} -- {A} quick and dirty usability scale.
\newblock In P.~W. Jordan, B.~Thomas, I.~L. McClelland, and B.~Weerdmeester,
  eds., {\em Usability Evaluation in Industry}. Taylor \& Francis, 1996.

\bibitem{calabrese2015urban}
F.~Calabrese, L.~Ferrari, and V.~D. Blondel.
\newblock Urban sensing using mobile phone network data: A survey of research.
\newblock {\em ACM Computing Surveys}, 47(2):25:1--25:20, 2015. doi: {{%
10\hspace{.1pt}\discretionary{.}{%
}{.}\hspace{.4pt}1145\discretionary{/}{%
}{/}2655691}}


\bibitem{chandler2015immersive}
T.~Chandler, M.~Cordeil, T.~Czauderna, T.~Dwyer, J.~Glowacki, C.~Goncu,
  M.~Klapperstueck, K.~Klein, K.~Marriott, F.~Schreiber, et~al.
\newblock Immersive analytics.
\newblock In {\em 2015 Big Data Visual Analytics (BDVA)}, pp. 1--8. IEEE,
  September 2015. doi: {{%
10\hspace{.1pt}\discretionary{.}{%
}{.}\hspace{.4pt}1109\discretionary{/}{%
}{/}BDVA\hspace{.1pt}\discretionary{.}{%
}{.}\hspace{.4pt}2015\hspace{.1pt}\discretionary{.}{%
}{.}\hspace{.4pt}7314296}}


\bibitem{chen2011exploratory}
J.~Chen, S.-L. Shaw, H.~Yu, F.~Lu, Y.~Chai, and Q.~Jia.
\newblock Exploratory data analysis of activity diary data: A space--time {GIS}
  approach.
\newblock {\em Journal of Transport Geography}, 19(3):394--404, 2011. doi: {{%
10\hspace{.1pt}\discretionary{.}{%
}{.}\hspace{.4pt}1016\discretionary{/}{%
}{/}j\hspace{.1pt}\discretionary{.}{%
}{.}\hspace{.4pt}jtrangeo\hspace{.1pt}\discretionary{.}{%
}{.}\hspace{.4pt}2010\hspace{.1pt}\discretionary{.}{%
}{.}\hspace{.4pt}11\hspace{.1pt}\discretionary{.}{%
}{.}\hspace{.4pt}002}}


\bibitem{cockburn2009review}
A.~Cockburn, A.~Karlson, and B.~B. Bederson.
\newblock A review of overview+detail, zooming, and focus+context interfaces.
\newblock {\em ACM Computing Surveys}, 41(1):2:1--2:31, Jan. 2009. doi: {{%
10\hspace{.1pt}\discretionary{.}{%
}{.}\hspace{.4pt}1145\discretionary{/}{%
}{/}1456650\hspace{.1pt}\discretionary{.}{%
}{.}\hspace{.4pt}1456652}}


\bibitem{coffey2011slice}
D.~Coffey, N.~Malbraaten, T.~Le, I.~Borazjani, F.~Sotiropoulos, and D.~F.
  Keefe.
\newblock Slice wim: A multi-surface, multi-touch interface for overview+detail
  exploration of volume datasets in virtual reality.
\newblock In {\em Symposium on Interactive 3D Graphics and Games}, pp.
  191--198. ACM, New York, NY, USA, 2011.

\bibitem{cordeil2017design}
M.~Cordeil, B.~Bach, Y.~Li, E.~Wilson, and T.~Dwyer.
\newblock Design space for spatio-data coordination: Tangible interaction
  devices for immersive information visualisation.
\newblock In {\em Proceedings of IEEE Pacific Visualization Symposium
  (PacificVis)}, pp. 46--50. IEEE, April 2017. doi: {{%
10\hspace{.1pt}\discretionary{.}{%
}{.}\hspace{.4pt}1109\discretionary{/}{%
}{/}PACIFICVIS\hspace{.1pt}\discretionary{.}{%
}{.}\hspace{.4pt}2017\hspace{.1pt}\discretionary{.}{%
}{.}\hspace{.4pt}8031578}}


\bibitem{demvsar2010space}
U.~Dem{\v{s}}ar and K.~Virrantaus.
\newblock Space--time density of trajectories: Exploring spatio-temporal
  patterns in movement data.
\newblock {\em International Journal of Geographical Information Science},
  24(10):1527--1542, 2010. doi: {{%
10\hspace{.1pt}\discretionary{.}{%
}{.}\hspace{.4pt}1080\discretionary{/}{%
}{/}13658816\hspace{.1pt}\discretionary{.}{%
}{.}\hspace{.4pt}2010\hspace{.1pt}\discretionary{.}{%
}{.}\hspace{.4pt}511223}}


\bibitem{drouhard2015immersive}
M.~Drouhard, C.~A. Steed, S.~Hahn, T.~Proffen, J.~Daniel, and M.~Matheson.
\newblock Immersive visualization for materials science data analysis using the
  oculus rift.
\newblock In {\em 2015 IEEE International Conference on Big Data (Big Data)},
  pp. 2453--2461. IEEE, October 2015. doi: {{%
10\hspace{.1pt}\discretionary{.}{%
}{.}\hspace{.4pt}1109\discretionary{/}{%
}{/}BigData\hspace{.1pt}\discretionary{.}{%
}{.}\hspace{.4pt}2015\hspace{.1pt}\discretionary{.}{%
}{.}\hspace{.4pt}7364040}}


\bibitem{wagnerfilho2019cga}
J.~A.~W. {Filho}, C.~M. D.~S. {Freitas}, and L.~{Nedel}.
\newblock Comfortable immersive analytics with the virtualdesk metaphor.
\newblock {\em IEEE Computer Graphics and Applications}, 39(3):41--53, May
  2019. doi: {{%
10\hspace{.1pt}\discretionary{.}{%
}{.}\hspace{.4pt}1109\discretionary{/}{%
}{/}MCG\hspace{.1pt}\discretionary{.}{%
}{.}\hspace{.4pt}2019\hspace{.1pt}\discretionary{.}{%
}{.}\hspace{.4pt}2898856}}


\bibitem{gatalsky2004interactive}
P.~Gatalsky, N.~Andrienko, and G.~Andrienko.
\newblock Interactive analysis of event data using space-time cube.
\newblock In {\em International Conference on Information Visualisation (IV)},
  pp. 145--152. IEEE, July 2004. doi: {{%
10\hspace{.1pt}\discretionary{.}{%
}{.}\hspace{.4pt}1109\discretionary{/}{%
}{/}IV\hspace{.1pt}\discretionary{.}{%
}{.}\hspace{.4pt}2004\hspace{.1pt}\discretionary{.}{%
}{.}\hspace{.4pt}1320137}}


\bibitem{gonccalves2015cartographic}
T.~Gon{\c{c}}alves, A.~P. Afonso, and B.~Martins.
\newblock Cartographic visualization of human trajectory data: Overview and
  analysis.
\newblock {\em Journal of Location Based Services}, 9(2):138--166, 2015. doi:
  {{%
10\hspace{.1pt}\discretionary{.}{%
}{.}\hspace{.4pt}1080\discretionary{/}{%
}{/}17489725\hspace{.1pt}\discretionary{.}{%
}{.}\hspace{.4pt}2015\hspace{.1pt}\discretionary{.}{%
}{.}\hspace{.4pt}1074736}}


\bibitem{gonccalves2016not}
T.~Gon\c{c}alves, A.~P. Afonso, and B.~Martins.
\newblock Why not both?: Combining {2D} maps and {3D} space-time cubes for
  human trajectory data visualization.
\newblock In {\em Proceedings of the 30th International BCS Human Computer
  Interaction Conference: Fusion!}, HCI '16, pp. 22:1--22:10. BCS Learning \&
  Development Ltd., Swindon, UK, 2016. doi: {{%
10\hspace{.1pt}\discretionary{.}{%
}{.}\hspace{.4pt}14236\discretionary{/}{%
}{/}ewic\discretionary{/}{%
}{/}HCI2016\hspace{.1pt}\discretionary{.}{%
}{.}\hspace{.4pt}22}}


\bibitem{hagerstraand1970people}
T.~H{\"a}gerstraand.
\newblock What about people in regional science?
\newblock {\em Papers in Regional Science}, 24(1):7--24, 1970.

\bibitem{hart2006nasa}
S.~G. Hart.
\newblock Nasa-task load index ({NASA-TLX}); 20 years later.
\newblock In {\em Proceedings of the Human Factors and Ergonomics Society
  Annual Meeting}, vol.~50, pp. 904--908. Sage Publications Sage CA: Los
  Angeles, CA, 2006.

\bibitem{hurter2018fiberclay}
C.~{Hurter}, N.~H. {Riche}, S.~M. {Drucker}, M.~{Cordeil}, R.~{Alligier}, and
  R.~{Vuillemot}.
\newblock Fiberclay: Sculpting three dimensional trajectories to reveal
  structural insights.
\newblock {\em IEEE Transactions on Visualization and Computer Graphics},
  25(1):704--714, Jan 2019. doi: {{%
10\hspace{.1pt}\discretionary{.}{%
}{.}\hspace{.4pt}1109\discretionary{/}{%
}{/}TVCG\hspace{.1pt}\discretionary{.}{%
}{.}\hspace{.4pt}2018\hspace{.1pt}\discretionary{.}{%
}{.}\hspace{.4pt}2865191}}


\bibitem{jankowski2015advances}
J.~Jankowski and M.~Hachet.
\newblock Advances in interaction with {3D} environments.
\newblock {\em Computer Graphics Forum}, 34(1):152--190, 2015. doi: {{%
10\hspace{.1pt}\discretionary{.}{%
}{.}\hspace{.4pt}1111\discretionary{/}{%
}{/}cgf\hspace{.1pt}\discretionary{.}{%
}{.}\hspace{.4pt}12466}}


\bibitem{geotime2004}
T.~{Kapler} and W.~{Wright}.
\newblock Geotime information visualization.
\newblock In {\em IEEE Symposium on Information Visualization}, pp. 25--32.
  IEEE, Oct 2004. doi: {{%
10\hspace{.1pt}\discretionary{.}{%
}{.}\hspace{.4pt}1109\discretionary{/}{%
}{/}INFVIS\hspace{.1pt}\discretionary{.}{%
}{.}\hspace{.4pt}2004\hspace{.1pt}\discretionary{.}{%
}{.}\hspace{.4pt}27}}


\bibitem{kennedy2003configural}
R.~S. Kennedy, J.~M. Drexler, D.~E. Compton, K.~M. Stanney, D.~S. Lanham, and
  D.~L. Harm.
\newblock Configural scoring of simulator sickness, cybersickness and space
  adaptation syndrome: Similarities and differences.
\newblock In L.~J. Hettinger and M.~W. Haas, eds., {\em Virtual and adaptive
  environments: Applications, implications, and human performance issues},
  chap.~12, pp. 247--276. CRC Press, 2003.

\bibitem{kjellin2010different}
A.~Kjellin, L.~W. Pettersson, S.~Seipel, and M.~Lind.
\newblock Different levels of {3D}: An evaluation of visualized discrete
  spatiotemporal data in space-time cubes.
\newblock {\em Information Visualization}, 9(2):152--164, 2010. doi: {{%
10\hspace{.1pt}\discretionary{.}{%
}{.}\hspace{.4pt}1057\discretionary{/}{%
}{/}ivs\hspace{.1pt}\discretionary{.}{%
}{.}\hspace{.4pt}2009\hspace{.1pt}\discretionary{.}{%
}{.}\hspace{.4pt}8}}


\bibitem{kjellin2010evaluating}
A.~Kjellin, L.~W. Pettersson, S.~Seipel, and M.~Lind.
\newblock Evaluating {2D} and {3D} visualizations of spatiotemporal
  information.
\newblock {\em ACM Transactions on Applied Perception (TAP)}, 7(3):19:1--19:23,
  2010. doi: {{%
10\hspace{.1pt}\discretionary{.}{%
}{.}\hspace{.4pt}1145\discretionary{/}{%
}{/}1773965\hspace{.1pt}\discretionary{.}{%
}{.}\hspace{.4pt}1773970}}


\bibitem{kraak2003space}
M.-J. Kraak.
\newblock The space-time cube revisited from a geovisualization perspective.
\newblock In {\em Proc. 21st International Cartographic Conference}, pp.
  1988--1996. Citeseer, 2003.

\bibitem{kraak2006visualization}
M.-J. Kraak.
\newblock Visualization viewpoints: Beyond geovisualization.
\newblock {\em IEEE Computer Graphics and Applications}, 26(4):6--9, July 2006.
  doi: {{%
10\hspace{.1pt}\discretionary{.}{%
}{.}\hspace{.4pt}1109\discretionary{/}{%
}{/}MCG\hspace{.1pt}\discretionary{.}{%
}{.}\hspace{.4pt}2006\hspace{.1pt}\discretionary{.}{%
}{.}\hspace{.4pt}74}}


\bibitem{kraak2008geovisualization}
M.-J. Kraak.
\newblock Geovisualization and time -- new opportunities for the space-time
  cube.
\newblock {\em Geographic visualization: Concepts, tools and applications}, pp.
  293--306, 2008. doi: {{%
10\hspace{.1pt}\discretionary{.}{%
}{.}\hspace{.4pt}1002\discretionary{/}{%
}{/}9780470987643\hspace{.1pt}\discretionary{.}{%
}{.}\hspace{.4pt}ch15}}


\bibitem{kristensson2009evaluation}
P.~O. Kristensson, N.~Dahlback, D.~Anundi, M.~Bjornstad, H.~Gillberg,
  J.~Haraldsson, I.~Martensson, M.~Nordvall, and J.~Stahl.
\newblock An evaluation of space time cube representation of spatiotemporal
  patterns.
\newblock {\em IEEE Transactions on Visualization and Computer Graphics},
  15(4):696--702, July 2009. doi: {{%
10\hspace{.1pt}\discretionary{.}{%
}{.}\hspace{.4pt}1109\discretionary{/}{%
}{/}TVCG\hspace{.1pt}\discretionary{.}{%
}{.}\hspace{.4pt}2008\hspace{.1pt}\discretionary{.}{%
}{.}\hspace{.4pt}194}}


\bibitem{kveladze2017researching}
I.~Kveladze, M.~Kraak, and C.~van Elzakker.
\newblock Researching the usability of a geovisual analytics environment for
  the exploration and analysis of different datasets.
\newblock In {\em The 28th International Cartographic Conference, 2-7 July
  2017, Washington DC, USA}, 2017.

\bibitem{kveladze2012we}
I.~Kveladze and M.-J. Kraak.
\newblock What do we know about the space-time cube from cartographic and
  usability perspective.
\newblock {\em Columbus, Ohio, USA: Proceedings of Autocarto}, pp. 16--18,
  2012.

\bibitem{kveladze2013methodological}
I.~Kveladze, M.-J. Kraak, and C.~P. van Elzakker.
\newblock A methodological framework for researching the usability of the
  space-time cube.
\newblock {\em The Cartographic Journal}, 50(3):201--210, 2013. doi: {{%
10\hspace{.1pt}\discretionary{.}{%
}{.}\hspace{.4pt}1179\discretionary{/}{%
}{/}1743277413Y\hspace{.1pt}\discretionary{.}{%
}{.}\hspace{.4pt}0000000061}}


\bibitem{kveladze2015space}
I.~Kveladze, M.-J. Kraak, and C.~P. Van~Elzakker.
\newblock The space-time cube as part of a geovisual analytics environment to
  support the understanding of movement data.
\newblock {\em International Journal of Geographical Information Science},
  29(11):2001--2016, 2015. doi: {{%
10\hspace{.1pt}\discretionary{.}{%
}{.}\hspace{.4pt}1080\discretionary{/}{%
}{/}13658816\hspace{.1pt}\discretionary{.}{%
}{.}\hspace{.4pt}2015\hspace{.1pt}\discretionary{.}{%
}{.}\hspace{.4pt}1058386}}


\bibitem{kveladze2018cartographic}
I.~Kveladze, M.-J. Kraak, and C.~P. van Elzakker.
\newblock Cartographic design and the space-time cube.
\newblock {\em The Cartographic Journal}, 56(1):73--90, 2019. doi: {{%
10\hspace{.1pt}\discretionary{.}{%
}{.}\hspace{.4pt}1080\discretionary{/}{%
}{/}00087041\hspace{.1pt}\discretionary{.}{%
}{.}\hspace{.4pt}2018\hspace{.1pt}\discretionary{.}{%
}{.}\hspace{.4pt}1495898}}


\bibitem{kveladze2012recognition}
I.~Kveladze, S.~van~der Spek, and M.-J. Kraak.
\newblock The recognition of temporal patterns in pedestrian behaviour using
  visual exploration tools.
\newblock In {\em 7th International Conference, GIScience 2012, Columbus, OH,
  USA, September, Proceedings 7478 (Paperback)}, 2012.

\bibitem{kwon2016study}
O.-H. Kwon, C.~Muelder, K.~Lee, and K.-L. Ma.
\newblock A study of layout, rendering, and interaction methods for immersive
  graph visualization.
\newblock {\em IEEE Transactions on Visualization and Computer Graphics},
  22(7):1802--1815, July 2016. doi: {{%
10\hspace{.1pt}\discretionary{.}{%
}{.}\hspace{.4pt}1109\discretionary{/}{%
}{/}TVCG\hspace{.1pt}\discretionary{.}{%
}{.}\hspace{.4pt}2016\hspace{.1pt}\discretionary{.}{%
}{.}\hspace{.4pt}2520921}}


\bibitem{li2010visual}
X.~Li, A.~{\c{C}}{\"o}ltekin, and M.-J. Kraak.
\newblock Visual exploration of eye movement data using the space-time-cube.
\newblock In {\em International Conference on Geographic Information Science},
  pp. 295--309. Springer, 2010. doi: {{%
10\hspace{.1pt}\discretionary{.}{%
}{.}\hspace{.4pt}1007\discretionary{/}{%
}{/}978\discretionary{%
}{-}{-}3\discretionary{%
}{-}{-}642\discretionary{%
}{-}{-}15300\discretionary{%
}{-}{-}6\_21}}


\bibitem{immersiveanalyticsbook}
K.~Marriott, F.~Schreiber, T.~Dwyer, K.~Klein, N.~H. Riche, T.~Itoh,
  W.~Stuerzlinger, and B.~H. Thomas, eds.
\newblock {\em Immersive Analytics}, vol. 11190 of {\em Lecture Notes in
  Computer Science}.
\newblock Springer, 2018. doi: {{%
10\hspace{.1pt}\discretionary{.}{%
}{.}\hspace{.4pt}1007\discretionary{/}{%
}{/}978\discretionary{%
}{-}{-}3\discretionary{%
}{-}{-}030\discretionary{%
}{-}{-}01388\discretionary{%
}{-}{-}2}}


\bibitem{mine1997moving}
M.~R. Mine, F.~P. Brooks~Jr, and C.~H. Sequin.
\newblock Moving objects in space: Exploiting proprioception in
  virtual-environment interaction.
\newblock In {\em Proceedings of the 24th Annual Conference on Computer
  Graphics and Interactive Techniques}, pp. 19--26. ACM Press/Addison-Wesley
  Publishing Co., New York, NY, USA, 1997. doi: {{%
10\hspace{.1pt}\discretionary{.}{%
}{.}\hspace{.4pt}1145\discretionary{/}{%
}{/}258734\hspace{.1pt}\discretionary{.}{%
}{.}\hspace{.4pt}258747}}


\bibitem{morgan2010visual}
J.~D. Morgan.
\newblock {\em A visual time-geographic approach to crime mapping}.
\newblock The Florida State University, 2010.

\bibitem{nguyen2017bees}
H.~Nguyen, F.~Wang, R.~Williams, U.~Engelke, A.~Kruger, and P.~d. Souza.
\newblock Immersive visual analysis of insect flight behaviour.
\newblock In {\em IEEE VIS Workshop on Immersive Analytics: Exploring Future
  Interaction and Visualization Technologies for Data Analytics}. IEEE, 2017.

\bibitem{noulas2011empirical}
A.~Noulas, S.~Scellato, C.~Mascolo, and M.~Pontil.
\newblock An empirical study of geographic user activity patterns in
  foursquare.
\newblock In {\em Fifth International AAAI Conference on Weblogs and Social
  Media}. AAAI, 2011.

\bibitem{okada2018}
K.~{Okada}, M.~{Yoshida}, T.~{Itoh}, T.~{Czauderna}, and K.~{Stephens}.
\newblock {VR} system for spatio-temporal visualization of tweet data.
\newblock In {\em International Conference Information Visualisation (IV)}, pp.
  91--95, July 2018. doi: {{%
10\hspace{.1pt}\discretionary{.}{%
}{.}\hspace{.4pt}1109\discretionary{/}{%
}{/}iV\hspace{.1pt}\discretionary{.}{%
}{.}\hspace{.4pt}2018\hspace{.1pt}\discretionary{.}{%
}{.}\hspace{.4pt}00026}}


\bibitem{pallant2013spss}
J.~Pallant.
\newblock {\em SPSS survival manual}.
\newblock McGraw-Hill Education (UK), 2013.

\bibitem{peuquet1994s}
D.~J. Peuquet.
\newblock It's about time: A conceptual framework for the representation of
  temporal dynamics in geographic information systems.
\newblock {\em Annals of the Association of American Geographers},
  84(3):441--461, 1994. doi: {{%
10\hspace{.1pt}\discretionary{.}{%
}{.}\hspace{.4pt}1111\discretionary{/}{%
}{/}j\hspace{.1pt}\discretionary{.}{%
}{.}\hspace{.4pt}1467\discretionary{%
}{-}{-}8306\hspace{.1pt}\discretionary{.}{%
}{.}\hspace{.4pt}1994\hspace{.1pt}\discretionary{.}{%
}{.}\hspace{.4pt}tb01869\hspace{.1pt}\discretionary{.}{%
}{.}\hspace{.4pt}x}}


\bibitem{saenz2017reexamining}
M.~Saenz, A.~Baigelenov, Y.-H. Hung, and P.~Parsons.
\newblock Reexamining the cognitive utility of {3D} visualizations using
  augmented reality holograms.
\newblock In {\em IEEE VIS Workshop on Immersive Analytics: Exploring Future
  Interaction and Visualization Technologies for Data Analytics}. IEEE, 2017.

\bibitem{satriadi2019freehand}
K.~A. Satriadi, B.~Ens, M.~Cordeil, T.~Czauderna, W.~Willett, and B.~Jenny.
\newblock Augmented reality map navigation with freehand gestures.
\newblock In {\em Proceedings of the 26th IEEE Conference on Virtual Reality
  and 3D User Interfaces}. IEEE, March 2019.

\bibitem{ssin2019geogate}
S.~Y. Ssin, J.~A. Walsh, R.~T. Smith, A.~Cunningham, and B.~H. Thomas.
\newblock Geogate: Correlating geo-temporal datasets using an augmented reality
  space-time cube and tangible interactions.
\newblock In {\em Proceedings of the 26th IEEE Conference on Virtual Reality
  and 3D User Interfaces}, March 2019.

\bibitem{theuns2017visualising}
J.~Theuns.
\newblock Visualising origin-destination data with virtual reality: Functional
  prototypes and a framework for continued {VR} research at the itc faculty.
\newblock {B.S.} thesis, University of Twente, 2017.

\bibitem{wagner2018virtualdesk}
J.~A. Wagner~Filho, C.~M. Freitas, and L.~Nedel.
\newblock {VirtualDesk: A Comfortable and Efficient Immersive Information
  Visualization Approach}.
\newblock {\em Computer Graphics Forum}, 37(3):415--426, 2018. doi: {{%
10\hspace{.1pt}\discretionary{.}{%
}{.}\hspace{.4pt}1111\discretionary{/}{%
}{/}cgf\hspace{.1pt}\discretionary{.}{%
}{.}\hspace{.4pt}13430}}


\bibitem{WagnerFilho2018VR}
J.~A. {Wagner Filho}, M.~F. Rey, C.~M. D.~S. Freitas, and L.~Nedel.
\newblock Immersive visualization of abstract information: An evaluation on
  dimensionally-reduced data scatterplots.
\newblock In {\em Proceedings of the 25th IEEE Conference on Virtual Reality
  and 3D User Interfaces}. IEEE, March 2018.

\bibitem{walsh2018tangible}
J.~Walsh, A.~Cunningham, R.~Smith, and B.~Thomas.
\newblock Tangible braille plot: Tangibly exploring geo-temporal data in
  virtual reality.
\newblock In {\em 2018 International Symposium on Big Data Visual and Immersive
  Analytics (BDVA)}, pp. 1--6. IEEE, 2018.

\bibitem{walsh2016braille}
J.~A. {Walsh}, J.~{Zucco}, R.~T. {Smith}, and B.~H. {Thomas}.
\newblock Temporal-geospatial cooperative visual analysis.
\newblock In {\em 2016 Big Data Visual Analytics (BDVA)}, pp. 1--8, Nov 2016.
  doi: {{%
10\hspace{.1pt}\discretionary{.}{%
}{.}\hspace{.4pt}1109\discretionary{/}{%
}{/}BDVA\hspace{.1pt}\discretionary{.}{%
}{.}\hspace{.4pt}2016\hspace{.1pt}\discretionary{.}{%
}{.}\hspace{.4pt}7787050}}


\bibitem{Ware:2004:IVP}
C.~Ware.
\newblock {\em Information Visualization: Perception for Design}.
\newblock Morgan Kaufmann Publishers Inc., San Francisco, CA, USA,
  2\textsuperscript{nd} ed., 2004. doi: {{%
10\hspace{.1pt}\discretionary{.}{%
}{.}\hspace{.4pt}1016\discretionary{/}{%
}{/}B978\discretionary{%
}{-}{-}155860819\discretionary{%
}{-}{-}1\discretionary{/}{%
}{/}50001\discretionary{%
}{-}{-}7}}


\bibitem{wilcoxon1945individual}
F.~Wilcoxon.
\newblock Individual comparisons by ranking methods.
\newblock {\em Biometrics Bulletin}, 1(6):80--83, 1945.

\bibitem{willems2011evaluation}
N.~Willems, H.~Van De~Wetering, and J.~J. Van~Wijk.
\newblock Evaluation of the visibility of vessel movement features in
  trajectory visualizations.
\newblock {\em Computer Graphics Forum}, 30(3):801--810, June 2011. doi: {{%
10\hspace{.1pt}\discretionary{.}{%
}{.}\hspace{.4pt}1111\discretionary{/}{%
}{/}j\hspace{.1pt}\discretionary{.}{%
}{.}\hspace{.4pt}1467\discretionary{%
}{-}{-}8659\hspace{.1pt}\discretionary{.}{%
}{.}\hspace{.4pt}2011\hspace{.1pt}\discretionary{.}{%
}{.}\hspace{.4pt}01929\hspace{.1pt}\discretionary{.}{%
}{.}\hspace{.4pt}x}}


\bibitem{yao2014oculus}
R.~Yao, T.~Heath, A.~Davies, T.~Forsyth, N.~Mitchell, and P.~Hoberman.
\newblock Oculus {VR} best practices guide.
\newblock {\em Oculus VR}, 2014.

\bibitem{zielasko2017remain}
D.~Zielasko, B.~Weyers, M.~Bellgardt, S.~Pick, A.~Meibner, T.~Vierjahn, and
  T.~W. Kuhlen.
\newblock Remain seated: Towards fully-immersive desktop {VR}.
\newblock In {\em IEEE 3rd Workshop on Everyday Virtual Reality (WEVR)}, pp.
  1--6. IEEE, March 2017. doi: {{%
10\hspace{.1pt}\discretionary{.}{%
}{.}\hspace{.4pt}1109\discretionary{/}{%
}{/}WEVR\hspace{.1pt}\discretionary{.}{%
}{.}\hspace{.4pt}2017\hspace{.1pt}\discretionary{.}{%
}{.}\hspace{.4pt}7957707}}


\end{thebibliography}
\end{document}